\documentclass{jfm}
\usepackage{graphicx}
\usepackage{epstopdf, epsfig}
\usepackage{lineno}
\usepackage{color}
\usepackage{amsmath}

\shorttitle{Turbulence suppression in pipe flow}
\shortauthor{Xu Liu and others}

\title{Turbulence suppression by streamwise-varying wall rotation in pipe flow}

\author{Xu Liu\aff{1},
  Hongbo Zhu\aff{1},
  Yan Bao\aff{1,2,3}
  \corresp{\email{ybao@sjtu.edu.cn}},
  Dai Zhou\aff{1,2,3}\\
 \and Zhaolong Han\aff{1,2,3}}

\affiliation{\aff{1}School of Naval Architecture, Ocean and Civil Engineering, Shanghai Jiao Tong University, Shanghai 200240,
People's Republic of China
\aff{2}State Key Laboratory of Ocean Engineering, Shanghai Jiao Tong University, Shanghai 200240, People's Republic of China
\aff{3}Key Laboratory of Hydrodynamics of Ministry of Education, Shanghai 200240, People's Republic of China
}

\begin{document}

\maketitle

\begin{abstract}
Direct numerical simulations (DNSs) of turbulent pipe flow subjected to streamwise-varying wall rotation are performed. This control method is able to significantly reduce the friction drag and even relaminarize the flow under certain control parameters at friction Reynolds number $Re_\tau$=$Ru_\tau$/$\nu$=180, where $R$ is the pipe radius, $u_\tau$ is the friction velocity and $\nu$ is the kinematic viscosity. Two control parameters, which are velocity amplitude and wavelength, are considered. The sensitivity of the control efficiency on wavelength is found to be greater than on the amplitude. An annular boundary layer, called Spatial Stokes Layer (SSL), is formed by the wall rotation. Based on the thickness of SSL, two types of drag reduction scenarios are identified. When the thickness is low, the SSL acts as a spacer layer, 
obstructing the impact of fluids against the wall and hence reducing the shear stress. The flow structures in the outer layer are less affected and are stretched in streamwise direction due to the increased velocity gradient. Within the SSL, the turbulence intensity diminishes dramatically. When the thickness is large, a streamwise wavy pattern of near-wall low-speed streaks is formed. The orientation of streaks is found to lag behind the direction of local mean velocity in phase and conform to the direction of local resultant shear stress. The streamwise scale of near-wall flow structures is significantly reduced, resulting in the disruption of downstream development of flow structures and hence leading to the drag reduction. Besides, it is found that it requires both large enough thickness of SSL and velocity amplitude to relaminarize the turbulence. The relaminarization mechanism is that the annular SSL can continuously absorb energy from wall-normal stress due to the rotation effect, thereby the turbulence self-sustaining process could not be maintained. For the relaminarization cases, the laminar state is stable to even extremely large perturbations, which possibly makes laminar state the only fixed point for the whole system.   
\end{abstract}


\section{Introduction}\label{introduction}
In pipeline transportation, turbulence causes significantly larger skin-friction drag than laminar flows, resulting in extra energy input in order to maintain the desired mass flow rate. Thus, for long-term consideration, even very limited reduction in friction drag can drastically lower the pollutant emission and bring about high economic benefits, which prompts researchers to develop efficient flow control methodology to improve the transport efficiency. 

Currently, most of the control methods can be classified into two categories: passive and active control, among which the latter is more promising since it can achieve higher drag reduction. One of the implementable active control method is wall-motion-based control. Inspired by the fact that turbulent boundary layer can be temporarily suppressed by the sudden spanwise pressure gradient, \cite{jung1992suppression} first demonstrated, via direct numerical simulation (DNS), that temporal wall oscillation in spanwise direction is able to achieve as much as 40\% drag reduction in turbulent channel flow at $Re_\tau$ =200, with the optimal non-dimensionalized oscillating period being $T^+$=100. Here, the quantities with superscript + denotes a value in wall units. These results were later confirmed numerically both in channel (\cite{baron1995turbulent,miyake1997mechanism,choi2002drag,quadrio2004critical}) and pipe flow  (\cite{quadrio2000numerical,nikitin2000mechanism,coxe2019vorticity,duggleby2007effect}) and experimentally in turbulent boundary layer (\cite{laadhari1994turbulence,skandaji1997etude,trujillo1997turbulent,di2002particle}). Further, \cite{quadrio2004critical} showed that a maximum net energy saving of 7.3\% can be achieved in turbulent channel flow of $Re_\tau$=200 as extra energy is needed to actuate the wall motion. They also successfully achieved the scaling of drag reduction in certain period range, making it possible to develop a model to predict the drag reduction. Later numerical (\cite{ricco2008wall,yao2019reynolds,yuan2019phase}) and experimental studies  (\cite{choi2001mechanism,choi2002near,ricco2004effects}), both in plane geometry, reported qualitatively similar drag reduction at similar Reynolds number. It is noted that in turbulent boundary layer, the maximum drag reduction is generally lower than that in channel (\cite{skote2012temporal,skote2019wall}). For a full collection of the drag reduction data in  different geometries, the readers can refer to the review by \cite{ricco2021review}.

Based on the fact that near-wall (say, below $y^+$=15) turbulence structures convect at a constant speed  of $U_c^+$=10 (\cite{kim1993propagation}), \cite{viotti2009streamwise} transformed the temporal oscillation into spatial oscillation, which is called standing wave oscillation. The close relation between the optimal oscillating wavelength and optimal oscillating period through $U_c^+$ suggests a strong analogy of these two control methods. Nevertheless, it is found that the spatial oscillation is more effective than temporal oscillation, that is, the former can achieve more drag reduction as compared to the latter (\cite{yakeno2009spatio}). \citet{skote2013comparison} made a comparison between spatial and temporal wall oscillation in turbulent boundary layer. It is shown that the streamwise gradient of mean spanwise velocity lower the production of spanwise Reynolds stress, leading to a lower magnitude of spanwise Reynolds stress than that in temporal oscillation case. However, an unambiguous and complete paradigm that explains the difference is still lacking.

Since the temporal and spatial wall oscillation can both achieve considerable drag reduction, it is natural to combine them to form the streamwise traveling-wave control, which was done by \cite{quadrio2009streamwise} in turbulent channel flow at $Re_\tau$ =200. In their study, a large number of cases were performed to obtain a drag reduction map by varying the oscillating frequency and spatial wavenumber. Both drag increasing and reducing regions in control parameters space were identified and drag increase occurs when the phase speed roughly coincides with the near-wall convection velocity $U_c^+$. Further, \cite{quadrio2011laminar} discussed the relationship between the thickness of generalized Stokes layer and drag reduction. Later, it is found, in turbulent channel flow, that the drag reduction continuously deteriorates as the Reynolds number increases (\cite{hurst2014effect,gatti2013performance,gatti2016reynolds}). Particularly, in addition to the variation of oscillating period and spatial wavenumber, \cite{gatti2016reynolds} also includes the variation of velocity amplitude, forming an extensive sweep of control parameters. It is shown that the effects of outer large-scale motions on drag reduction is subordinate. The experiments considering streamwise traveling-wave control were conducted by \cite{auteri2010experimental} in pipe flow and \cite{bird2018experimental} in turbulent boundary layer, both confirming the DNS results of \cite{quadrio2009streamwise}. More recently, a drag reduction level of 25\% at $Re_\tau$=6000 and 13\% at $Re_\tau$=12800 are reported by \cite{marusic2021energy}, demonstrating the potential application prospects in practical problems.

It is certain that the success of spanwise wall oscillation in reducing the skin-friction drag is rooted in the disruption of self-sustaining process of near-wall turbulence, which involves the quasi-organized low/high-speed streaks and streamwise vortices. When temporal oscillation is imposed, the wall motion drags the near-wall streaks laterally, while the overriding streamwise vortices remain almost unaffected, causing the relative displacement between these two structures and hence their spatial coherence is disrupted  (\cite{akhavan1993control,ricco2004effects}). As a consequence, the intensity and frequency of burst-sweep activity is significantly attenuated and the friction drag is reduced. Besides, the near-wall high-speed streak is driven by the near-wall shear layer to intrude below the adjacent low-speed streak when the wall velocity is accelerating (\cite{quadrio2000numerical}) and the low-speed streak is significantly suppressed when the direction of streamwise vortex rotation counteracts the wall motion (\cite{choi2002drag}). Moreover, \cite{yakeno2014modification} observed that the inclination of quasi-streamwise vortices leads to a strong streamwise stretching, hence strengthening the energy exchange between normal stress components and thereby enhancing the rotational motions. This is in accordance with \cite{choi2002near} and \cite{choi2001mechanism}, who showed that a net negative spanwise vorticity is generated by the interaction of vortex sheet generated by Stokes layer with longitudinal vortices, resulting in the distortion of near-wall mean velocity profile. However, this scenario is doubted by \cite{touber2012near}, who claimed that the visualized streak inclination follows from the direction of shear strain rather than the wall velocity, in which the latter seems to be the basis on which the above scenario rests. The time behaviour of near-wall vortices and streaks are examined by \cite{zhou2006mechanism}. And \cite{yang2019modulation} studied the drag reduction deterioration by examining the impact of logarithmic attached coherent state.

Except for the studies that focus on the modification of near-wall turbulence structures, many studies, covering many other aspects, has been conducted to comprehensively investigate the effects of temporal wall oscillation. The initial response of fluids to wall motion is studied by \cite{quadrio2003initial} and \cite{xu2005transient} in channel flow, in which the longitudinal flow is found to cost longer time to reach its statistically stable state than spanwise flow and the suppression of inter-component exchange of turbulent kinetic energy (TKE) is responsible for the drag reduction. In the physical constraint of constant pressure gradient (CPG), \cite{ricco2012changes} showed that the enhancement of turbulent dissipation in absolute term is responsible for the initial decline of TKE and hence the increase of flow rate. This paradigm is corroborated by \cite{ge2017response}, who further links the subsequent attenuation of dissipation to the variation of vorticity component. However, in the condition of constant flow rate (CFR), \cite{agostini2014spanwise} concluded that dissipation rises when the drag increases and falls when the drag decreases by examining the phasewise variation of turbulence quantities. Similar results have also been reported by \cite{touber2012near} and \cite{yuan2019phase}. This contradiction stems from the fact that in CPG condition, the drag reduction manifests itself as the the increase of flow rate. Hence, for the CPG condition, comparison should be made between cases of controlled and uncontrolled flow with the same flow rate. In this sense, such contradiction disappears. Moreover, the effects of streamwise local-implemented wall oscillation on the downstream development of the flow are investigated by \cite{yudhistira2011direct} and \cite{skote2015drag} in turbulent boundary layer, and a spanwise locally-implemented wall oscillation in duct is studied by \cite{straub2017turbulent}.

Possibly due to similarity between temporal and spatial oscillation, relatively less attention is payed to flow modification by spatial wall oscillation. \cite{skote2011turbulent} applied spatial wall oscillation to turbulent boundary layer and observed that the maximum drag reduction occurs at the location of maximum wall velocity while the minimum corresponds to zero wall velocity. \cite{negi2015dns} did innovative research on the modulation of a single low-speed streak by spatial wall oscillation in laminar boundary layer. Lower wavenumber is observed to possess stronger suppression on velocity fluctuation in the laminar regime. However, when the flow undergoes transition, the mechanism is significantly different due to the nonlinear amplification process and hence the deduction for laminar regime is inapplicable. \cite{yakeno2009spatio} compared the temporal and spatial oscillation and found that the phase-dependency of Reynolds shear stress is more significant in spatial oscillation.

When such wall-motion control applies to turbulent pipe flow, the crucial difference, compared with plane geometry, is that relaminarization is more likely to occur at low Reynolds number. With temporal wall oscillation imposed, \cite{choi2002drag} showed that relaminarization occurs in pipe flow of $Re_\tau$=150 when oscillating period $T^+$$\textgreater$150 with velocity amplitude $A^+$=10 and $T^+$$\textgreater$100 with $A^+$=20 ($A$ is the velocity amplitude), while no relaminarization occurs in channel at even a lower Reynolds number $Re_\tau$ =100 with the same control parameter. \cite{nikitin2000mechanism} also reported relaminarization at $Re_\tau$=133 with ($A^+$,$T^+$)=(9,105) in pipe flow. Besides, the control with large oscillating period does not cause a drag increase in pipe flow, which is opposite to the channel-flow results  (\cite{jung1992suppression}). They attributed this discrepancy to the centrifugal forces in cylindrical geometry but no further details are provided. \cite{biggi2013riduzione} and \cite{xie2014turbulence} applied streamwise traveling-wave control to turbulent pipe flow at $Re_\tau$=200, with the former under the physical constraint of CFR and the latter under constant power input (CPI). They both observed relaminarization which does not occur in channel at the same Reynolds number and control parameters (\cite{quadrio2009streamwise}). \cite{quadrio2009streamwise} also reported relaminarization by streamwise traveling-wave in channel flow at $Re_\tau$=100. Furthermore, \cite{ming2019drag} focused on the drag increase of external flow along a cylinder subjected to wall rotation. They found that the rotation-induced vortices, which are the consequences of centrifugal effect, contribute dominantly to the drag increase. Hence, it is clear that the geometry difference can cause drastically different behaviour of flow when imposing the wall-motion control technique. 

To date, it is known that the thickness of Stokes layer is crucial to the drag reduction performance (\citet{quadrio2011laminar}) and most of studies are focusing on the quantitative relationship between thickness and drag reduction. But the detailed paradigm that explains how the thickness of Stokes layer affects the drag reduction and how the flow structures reacts to different thickness of Stokes layer is still lacking. Besides, much of the attention has been paid to drag reduction in planar flows while the number of studies focusing on pipe flow forced by wall motion (especially for the spatially non-uniform form) are far less. More important is that the geometry can cause significant difference, that is, under the same control parameters, relaminarization occurs in pipe, whereas in channel, the flow remains turbulent. The mechanism that causes such significant difference is still unclear (\citet{xie2014turbulence}). Hence, in present study, we consider the turbulent pipe flow with standing-wave control imposed to examine the drag reduction behaviour with the aid of DNS, in the hope of extending the wall-motion control database, providing a more detailed drag reduction scenario for different thickness of Stokes layer and gaining further insight into the physical mechanisms of relaminarization in pipe. A relatively small range of control parameter-sweep is conducted and the changes of flow field are shown by examining the turbulence statistics and time evolution of instantaneous flow fields. This paper is organized as follows. The computational details and control parameters are introduced in \S\ref{Methodology}. The drag reduction results and its comparison with previous studies and the energetic performance are given in \S\ref{Control Results}. The basic turbulence statistics and the modification of flow structures under control for the non-relaminarization cases are discussed in \S\ref{turbulent regime}. The transient dynamics and the physical mechanisms during the relaminarization process in pipe, the comparison with channel flow are explored in \S\ref{relaminarization}. Finally, \S\ref{summary} summarizes the main finding of this paper.
 
\section{Methodology}\label{Methodology}

\subsection{Computational details}\label{computational details}
\begin{figure}
\includegraphics[scale=0.8]{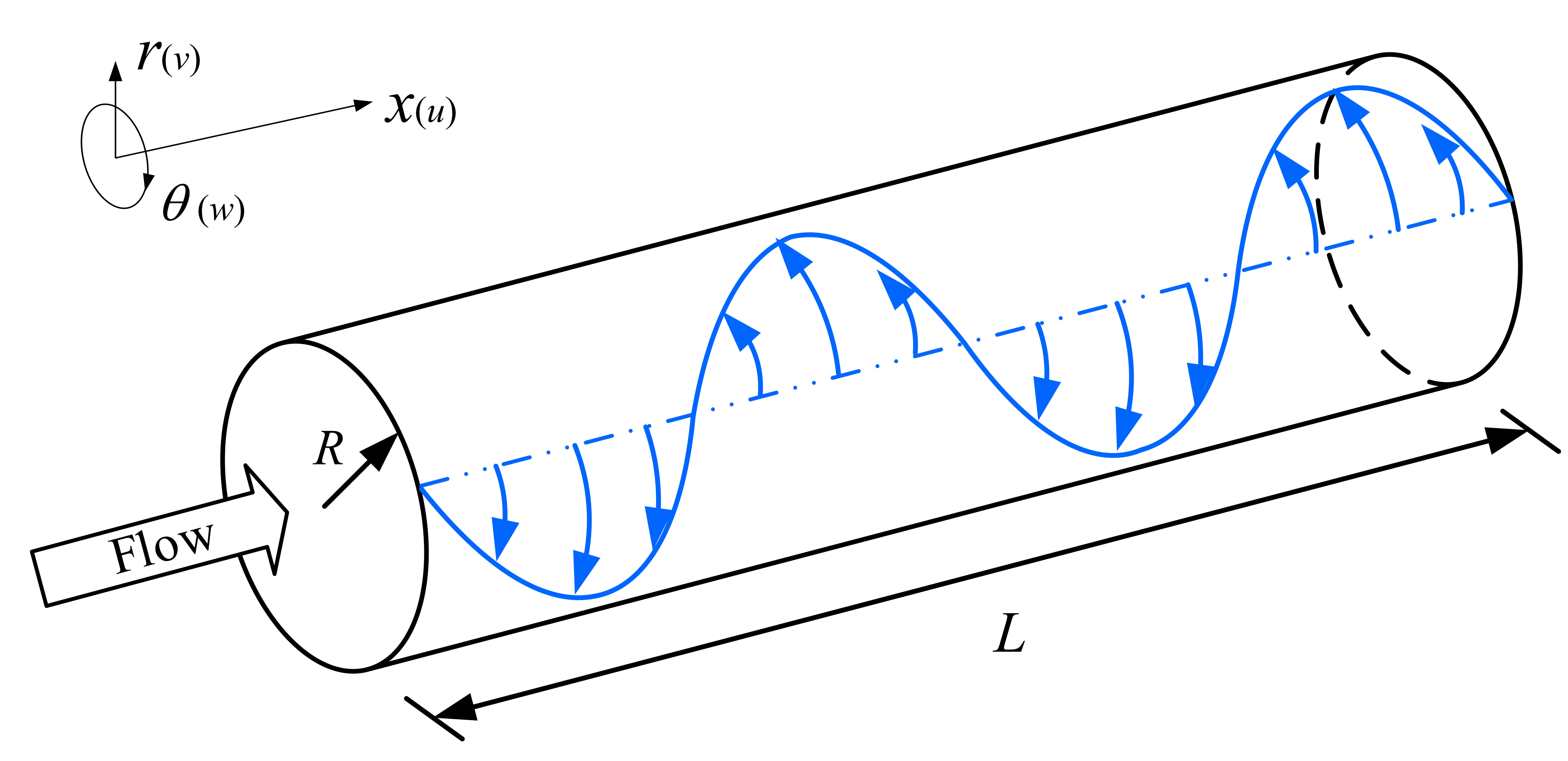}
\centering
\caption{\label{fig:computational domain} Schematic diagram of circular pipe with standing wave wall oscillation imposed. The length of arrow qualitatively denotes the magnitude of wall velocity. A pipe length of $L$=6$\pi$$R$ is employed. The corresponding velocity components are listed in parentheses.}
\end{figure}

We employ a cylindrical-coordinate spectral element-Fourier DNS solver \emph{Semtex}  (\cite{blackburn2019semtex,blackburn2004formulation}) to conduct the simulations at a friction Reynolds number $Re_\tau$=$u_\tau R/\nu$=180, where $u_\tau$ is the friction velocity, $R$ is the pipe radius and $\nu$ is the kinematic viscosity (based on bulk velocity $U_b$, it is approximately $Re_b$=$U_b R/\nu$=5300). This Reynolds number is consistent with those used in previous pipe-flow simulations (\cite{wu2008direct,eggels1994fully}). The physical model and computational domain are depicted in figure \ref{fig:computational domain}. No-slip and no-penetration boundary condition are applied at the pipe wall. The azimuthal component of wall velocity is:
\begin{equation}\label{equ:bc}
W=Asin(kx),
\end{equation}
where $A$ and $k$=2$\pi$/$\lambda$ are the amplitude and streamwise wavenumber respectively ($\lambda$ is the control wavelength). That means, the wall rotation velocity varies sinusoidally in streamwise direction but remains constant along azimuthal direction, implying azimuthal homogeneity.

The flow is driven by constant streamwise pressure gradient, yielding a constant friction velocity $u_\tau$ due to the balance between streamwise pressure gradient and wall shear stress, which allows unique wall-unit scaling. The flow starts from a laminar state, in which random perturbations are added to initiate the transition to turbulence. After the turbulence reaches a statistical steady state, a single field is selected as the starting point for the simulations of controlled cases with different combinations of $A$ and $\lambda$. During the whole process, the streamwise pressure gradient remains unchanged, which means that the drag reduction manifests itself as the variation of flow rate. The superscript + denotes normalization with respect to $u_\tau$=$\sqrt{\tau_w/\rho}$ for velocity and $\nu$/$u_\tau$ for distance, where $\tau_w$ is the wall shear stress and $\rho$ is the density of the fluid. Turbulence statistics were collected for more than 5500 viscous time units, which allows the fluids to travel more than 23 times through the pipe axial dimension at bulk velocity.

\begin{figure}
\centering
\includegraphics[scale=0.52]{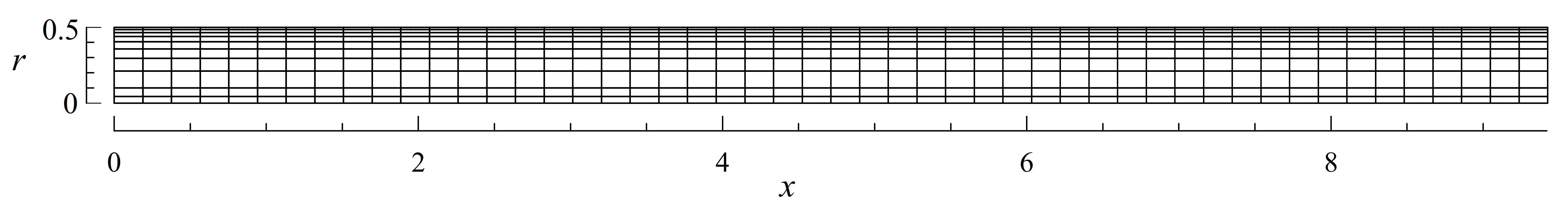}
\caption{\label{fig:mesh}Spectral element meshes for pipe, with 500 elements in the meridional semi-plane.}
\end{figure}

Regarding the pipe length, \cite{chin2010influence} suggested that a periodic pipe length of 8$\pi$$R$ seems to be sufficient to ensure that all statistics are not affected by the finite computational domain for $Re_\tau$=170-500. Based on the fact that the scale of large-scale motions (LSMs) in turbulent pipe flow ranges from 8$R$ to 16$R$ (\cite{kim1999very,guala2006large,morrison2004scaling}), \cite{wu2008direct} employed a pipe length of 15$R$ at $Re_b$=5300 and 44000 and demonstrated its adequacy by computing the two-point correlation data. The pipe length used in \cite{eggels1994fully} is 10$R$, which is proved to be too short to let the large-scale motion decouple from the computational domain. Hence, in consideration of computational costs, a periodic pipe length of 6$\pi$$R$ is chosen for current study.

The two-dimensional spectral element mesh, shown in figure \ref{fig:mesh}, is deployed to discretize the meridional semi-plane, together with Fourier expansions in azimuthal direction to represent the three-dimensional computational domain. The mesh consists of 50$\times$10 array of elements, in which the heights of the first two elements from pipe centerline is set equally and small enough to ensure the computational stability due to polar singularity at the pipe axis while the remaining element heights follows a geometric degression to the wall. We employ 10th-order nodal shape function (11 points along the edge of an element ($P$=11)), resulting in a minimum grid resolution of $\Delta$$r^+$=0.53 at the wall and a maximum of $\Delta$$r^+$=3.6 at the pipe centerline. The streamwise grid spacing is $\Delta$$x^+$=6.8. In the azimuthal direction, a number of 192 planes ($N_z$=192) are employed, yielding an azimuthal spacing of $\Delta$$(r\theta)^+$=5.9. The total computational nodes is approximately 1.2$\times$$10^7$. $x$,$r$,$\theta$,$t$ denotes the streamwise, radial, azimuthal direction and time, and the corresponding velocity components are $u$, $v$ and $w$ respectively. The wall-normal distance is defined as $y$=$R$-$r$. The unit length scale is pipe diameter $D$=2$R$. Moreover, the abbreviation ’SW’ used in this paper denotes ’standing wave’ control.

There are two main aspects discussed in this paper, one is the turbulence statistics for non-relaminarization cases (\S \ref{turbulent regime}) and the other is the transient dynamics for relaminarization cases (\S \ref{relaminarization}). In \S 4, figure \ref{fig:validation}(\emph{a}) and figure \ref{fig:channel_validation}(\emph{a}), a quantity $f$ averaged in $\theta$ and $t$ is denoted as $\widetilde{f}$ (also called phase-averaged quantity). A further average in $x$ of $\widetilde{f}$ is denoted as $\overline{f}$. A global quantity is defined as:
\begin{equation}\label{global quantity}
[f] = \int_0^R {\overline f } rdr
\end{equation}
The fluctuation velocity $\textbf{\emph{u}}'$ is calculated by the instantaneous value $\textbf{\emph{u}}$ subtracting its time-averaged value. In \S \ref{relaminarization}, figure \ref{fig:validation}(\emph{b}) and figure \ref{fig:channel_validation}(\emph{b}), for the above average operators, only time average is excluded. So the velocity fluctuations $\emph{\textbf{u}}'$ are calculated by instantaneous value $\emph{\textbf{u}}$ subtracting its circumferential (spanwise for channel)-averaged value $\widetilde{\emph{\textbf{u}}}$ since attentions are drawn to the time evolution of flow fields.

\subsection{ Control parameters}\label{control parameter}

\begin{table}
\centering
\begin{tabular}{cccc}
Case&Wavelength($\lambda^+$)&Amplitude($A^+$)&Drag reduction($D$)\\
\hline
0 & $\infty$ & 0 & -\\
1 & 565 & 12 & 42.14\% \\
2 & 1695 & 12 & RLM\\
3 & 3390 & 12 & RLM\\
4 & 1695 & 6 & 28.06\% \\
5 & 1695 & 30 & RLM\\
6 & $\infty$ & 12 & RLM\\
\end{tabular}
\caption{\label{tab:table1}Cases for different control parameters. Case 0 is the uncontrolled pipe. RLM denotes relaminarization, corresponding to a drag reduction value of $D$=0.89.}
\end{table}

In present study, we have explored the drag reduction performance of the standing wave wall oscillation in pipe flow by computing the variation of friction coefficient between the reference simulation and 6 additional computational cases, in which the wavelength $\lambda$ and velocity amplitude $A$ are varied independently.  The whole set of simulations are documented in table \ref{tab:table1}. 

Under the physical constraint of CPG, the drag reduction is defined as the ratio of the variation of friction coefficient $C_f$=2$\tau_w$/$U_b^2$ to the uncontrolled value $C_{f,0}$ (\cite{kasagi2009toward,ricco2012changes}), which can be written as:
\begin{equation}\label{drag reduction}
D = \frac{{{C_{f,0}} - {C_f}}}{{{C_{f,0}}}} = \frac{{U_b^2 - U_{b,0}^2}}{{U_b^2}}
\end{equation}

\subsection{ Validation}\label{validation}

\begin{figure}
\includegraphics[scale=0.85]{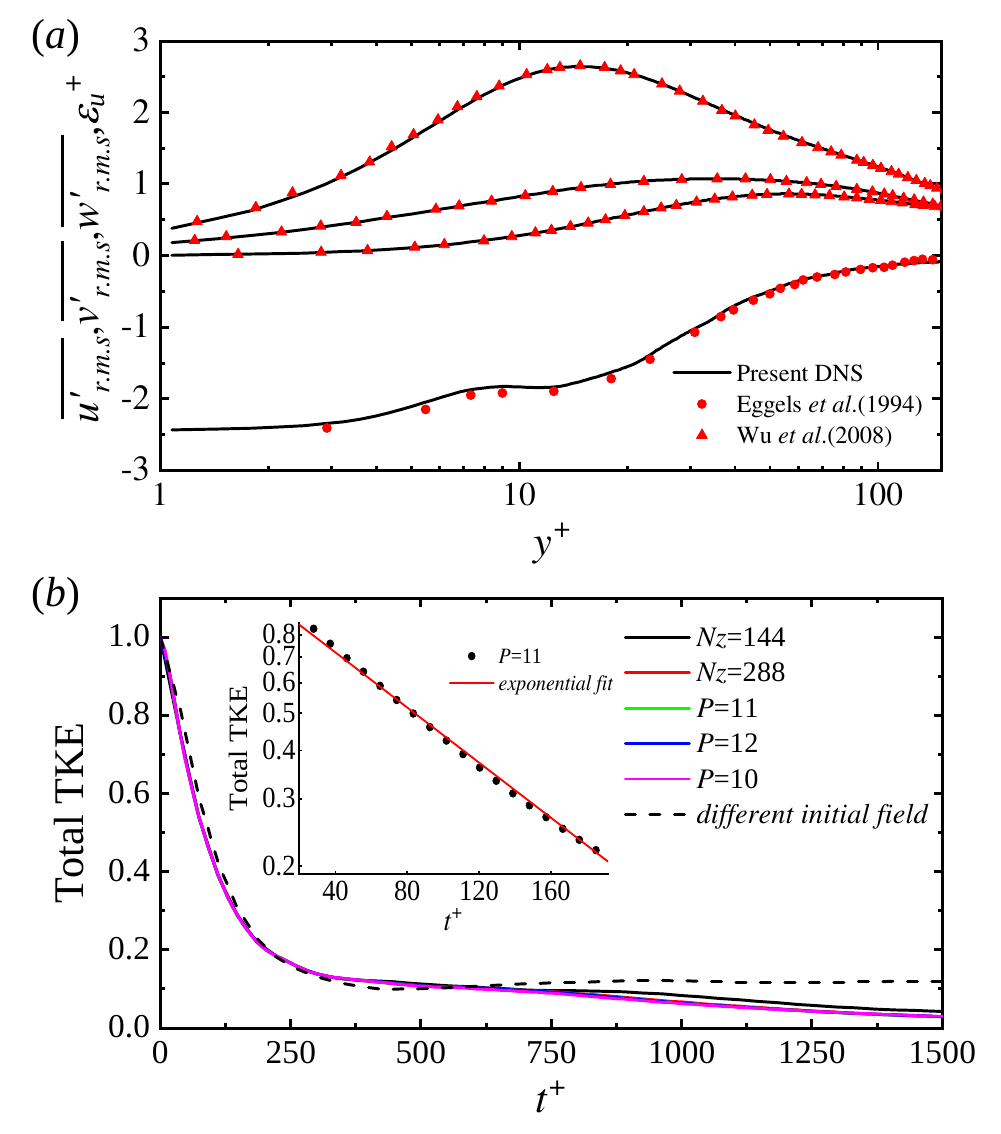}
\centering
\caption{\label{fig:validation} (\emph{a}) Comparison of turbulent statistics between present DNS results for uncontrolled case and previous literature. The dissipation rate of streamwise stress is scaled by $u_\tau^3/D$ and further divided by a scale factor of 36. (\emph{b}) Time evolution of total TKE ($\int_0^{2\pi}$[TKE]$d\theta$) during the relaminarization process for different meshes. The value of total TKE is normalized by the average value of uncontrolled case. The baseline mesh is $P=11$, $N_z$=192. For other four meshes, only one parameter (either $P$ or $N_z$) is varied to compare with the baseline mesh. The black dashed line represents the total TKE evolution for different initial field. The inset in (\emph{b}) display an exponential fit to the TKE evolution data. }
\end{figure}

Figure \ref{fig:validation}(\emph{a}) shows the comparison of turbulence intensity in uncontrolled pipe obtained by our DNS with that from previous pipe-flow database (\cite{wu2008direct,eggels1994fully}). Good agreement can be found for all statistics, including the dissipation rate of streamwise Reynolds stress, demonstrating the quantitative reliability of present simulations of fully-developed turbulent pipe flow.

On the other hand, in order to verify that the relaminarization is a physically-relevant process rather than consequence of insufficient mesh resolution, case 5 has been chosen for additional mesh resolution check. Such choice is based on the fact that $A^+$=30 is the maximum velocity amplitude employed in present study. Hence, the success of mesh independence validation of this extreme case allows us to be confident with the reliability of DNS data for other cases. Since relaminarization is a gradual process (\cite{sreenivasan1982laminarescent}) and it takes extremely long time for the flow to transform from turbulent state to fully-developed laminar Hagen-Poiseuille profile, it is reasonable that relaminarization occurs when the total turbulent kinetic energy (TKE) decreases to zero (\cite{lieu2010controlling}). Figure \ref{fig:validation}(\emph{b}) shows the time evolution of total TKE for case 5. Except for the baseline simulation, results from four additional meshes, for which either order of shape function ($P$-1) or number of Fourier planes ($N_z$) is varied with respect to baseline mesh, are also included. The collapse of all TKE evolution curves indicates that the present mesh is adequate and the relaminarization is indeed a physical process. Moreover, the complete flow excursion of relaminarization is calculated only for baseline mesh due to the extremely long computing time. No instability occurs and the flow indeed reaches the final laminar state. We also
change the initial flow field at which control is imposed for case 5, relaminarization also occurs with the same
pattern (black dashed line in figure \ref{fig:validation})(\emph{b}). Such independence from initial condition suggests that the relaminarization in pipe is robust.

\section{Control Results}\label{Control Results}

\subsection{\label{Drag reduction} Drag reduction results}

\begin{figure}
\includegraphics[scale=0.8]{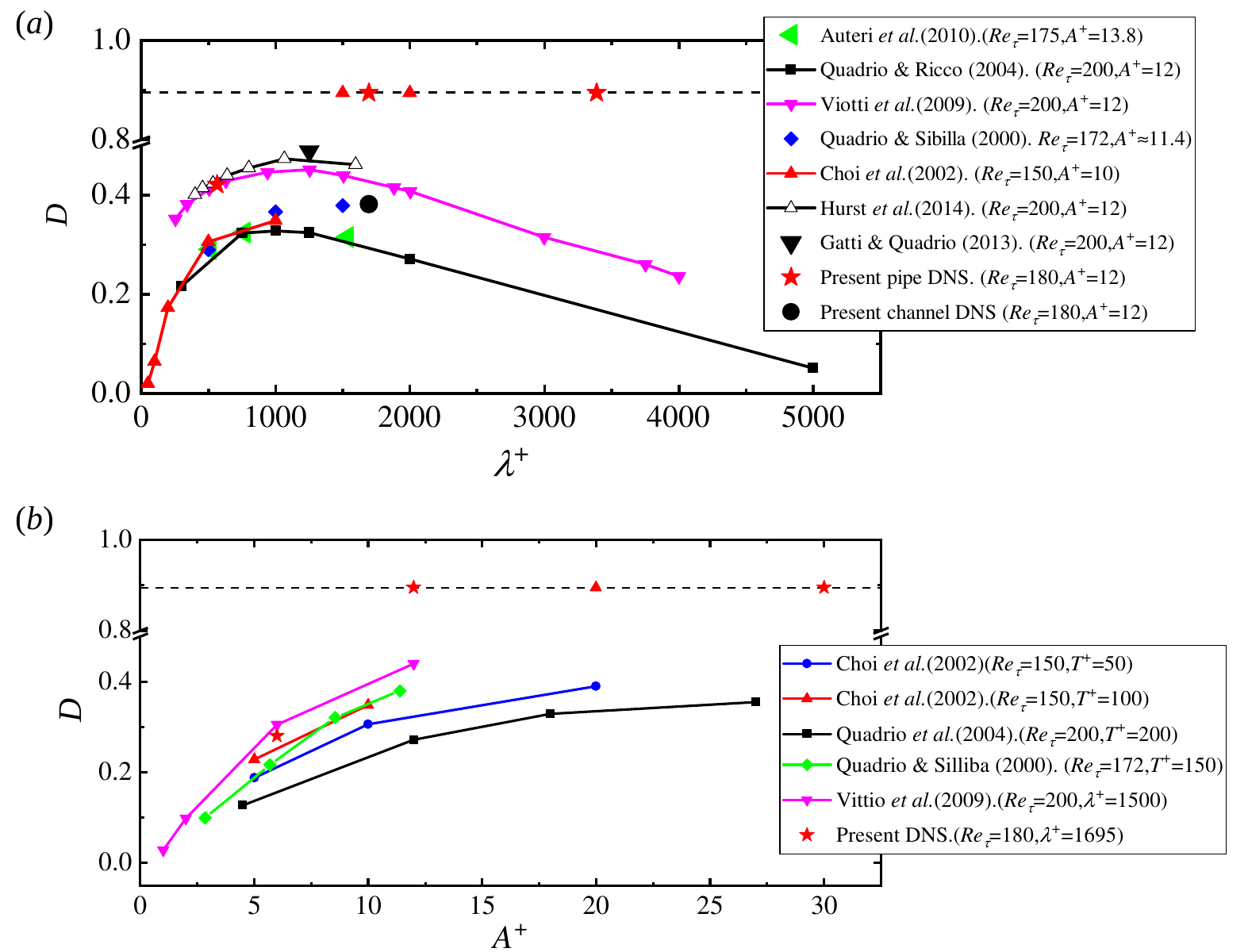}
\centering
\caption{\label{fig:drag reduction result}(\emph{a}) Variation of drag reduction with respect to wavelength at fixed amplitude. (\emph{b}) Variation of drag reduction with respect to amplitude at fixed wavelength. Data from previous literature are included for comparison. For temporal oscillation cases, the oscillating period is converted to wavelength via equation (\ref{equ:convert}). Dashed line represents the  drag reduction for laminar flow. Note that \citet{choi2002drag} did not provide the specific drag reduction value for relaminarization cases, thus here we simply put them on the dashed line. }
\end{figure}

The drag reduction results for different cases are shown in table \ref{tab:table1}. Relaminarization occurs for case 2,3 and 5, for which both the wavelength ($\lambda$) and amplitude ($A$) are relatively large. For non-relaminarization cases (case 1 and 4), the drag reduction is 42.14\% and 28.06\% respectively. It can be preliminarily observed that increasing the wavelength or amplitude can both leads to relaminarization, a trend which, however, will not occur in channel flow (\cite{viotti2009streamwise}). Hence, it is necessary to compare the present results with the existing literature data, as shown by figure \ref{fig:drag reduction result}. Also included in figure \ref{fig:drag reduction result} are the results from temporal wall oscillation in pipe (\cite{quadrio2000numerical,choi2002drag}) and channel flow (\cite{quadrio2004critical}) since the wavelength ($\lambda$) and period ($T$) can be linked by (\cite{viotti2009streamwise,quadrio2009streamwise})
\begin{equation}
\label{equ:convert}
{T^ + } = \frac{{{\lambda ^ + }}}{{{U_c^ + }}}.
\end{equation}

Figure \ref{fig:drag reduction result}(\emph{a}) presents the variation of drag reduction with respect to wavelength at fixed amplitude. In channel flow, with the increasing of wavelength, the drag reduction increases initially and then decreases, yielding an optimal wavelength of $\lambda^+$=1000-1250. For pipe flow, the result of low-wavelength case (case 1) coincides with that in channel (\cite{viotti2009streamwise}) while further increase in wavelength leads to relaminarization. Note that the simulations in \cite{viotti2009streamwise} is conducted in the constraint of CFR at $Re_\tau$=200, a drag reduction leads to a lower $Re_\tau$ which is hence comparable to present study. In present study, the finite pipe length restricts the maximum wavelength to 3390. But if the wavelength increases to infinity, it can be theoretically considered as a purely rotating pipe with circumferential wall velocity of $A^+$=12 in finite pipe domain. So we also performed a DNS of purely rotating pipe with $A^+$=12, whose results show that the turbulence also relaminarizes. Although the purely rotating wall can also relaminarize the pipe flow, it is evident that it requires more energy input since the wall rotates integrally at constant velocity, indicating the superiority of spatial wall oscillation control.

When considering the temporal wall oscillation, the same trend can be observed for channel with the optimal oscillating period located at $T^+$=100-125. It is already known that spatial oscillation is more effective in reducing the friction drag than temporal oscillation (\cite{viotti2009streamwise}). However, for pipe flow, the increase of oscillating period also leads to relaminarization (\cite{choi2002drag}) (For cases in \cite{quadrio2000numerical}, the oscillating period is not large enough to produce relaminarization). Besides, for low oscillating periods, the results from pipe and channel agrees well. Therefore, it can be concluded, at this relatively low Reynolds number, that for low wavelength or low period, there is no remarkable difference between pipe and channel. But for large wavelength and period, the turbulence relaminarizes in pipe, while in channel, the flow remains turbulent and the drag reduction deteriorates.

Part of the experimental data that pertains to standing wave control from \cite{auteri2010experimental}, in which streamwise traveling-wave are imposed at turbulent pipe flow, are also included for comparison. The lower drag reduction compared to numerical results can be attributed to the spatial transient at the upstream end of the actuated section (the section that the wall is rotating), as explained in \cite{auteri2010experimental}. Another issue that deserves to be noted is that no relaminarization is reported in experiment despite the fact that relaminarization is also achievable under the streamwise traveling-wave control (\cite{xie2014turbulence,biggi2013riduzione}). This is predictable, however. Although the laminar Hagen-Poiseuille flow in pipe is linear stable to infinitesimal disturbance for all Reynolds numbers, the perturbations from the discrete actuated section and the unsmoothness of pipe surface in real world are large enough to trigger the transition to turbulence at Reynolds number considered here and hence laminar state could not be maintained.  

Figure \ref{fig:drag reduction result}(\emph{b}) shows the variation of drag reduction as a function of amplitude at fixed wavelength, together with the data from literature for comparison. Again, drag reduction result of case 4 agrees well with that in channel \cite{viotti2009streamwise}, indicating strong similarity between pipe and channel for low-amplitude control. Similarly, relaminarization occurs at large amplitude for pipe flow. For channel flow, the literature data at high amplitude for standing wave control is very few and hence it is uncertain whether it relamainarizes at high amplitude. Nevertheless, the temporal oscillation data from \citet{quadrio2004critical} clearly shows that there is a saturation of drag reduction up to $A^+$=27. Also, according to \citet{choi2002drag}, the drag reduction shows a trend of saturation for $T^+$=50 but relaminarization occurs at larger amplitude for $T^+$=100. Since the thickness of temporal or spatial Stokes layer is closely related to oscillating period or wavelength (\cite{viotti2009streamwise}), it can be inferred that sufficiently large thickness of Stokes layer and simultaneously large amplitude are required for relaminarization in pipe flow.

\subsection{Energetic performance}\label{energetic performance}

It is obvious that the control law (\ref{equ:bc}) requires external energy to actuate the wall motion, which means additional energy are injected into the system. In this section, we evaluate whether a net energy saving could be achieved for this active control strategy in order to verify its possibility in real application. 

Multiplying equation (\ref{equ:spanwise momentum}) by $\widetilde{w}$ and taking average in $x$ direction, all the $x$-derivative terms vanish due to the streamwise periodicity of control law (\ref{equ:bc}). After that, by integrating in radial direction ($\int_0^R rdr$), we have the global energy balance for circumferential mean flow in pipe: 

\begin{equation}\label{equ:cir_mean_flow}
\left[ { - \widetilde {v'w'}\frac{{\partial \widetilde w}}{{\partial r}}} \right] + \left[ { - \widetilde {u'w'}\frac{{\partial \widetilde w}}{{\partial x}}} \right] + \underbrace{\left[ {\frac{{\widetilde {v'w'}\widetilde w}}{r}} \right]}_\textsc{$\wp$} \\
= \frac{1}{2}\nu [\triangledown ^2 {\widetilde w^2}] - \nu \left[ \left( \left({\frac{{{{\widetilde w}}}}{{{r}}}}\right)^2+{\left(\frac{{\partial \widetilde w}}{{\partial r}}\right)^2} + {\left(\frac{{\partial \widetilde w}}{{\partial x}}\right)^2}\right)\right]
\end{equation}

In the derivation of equation (\ref{equ:cir_mean_flow}), transport term disappears since Reynolds stress and velocity are zero at the wall. The three terms on the left-hand side represents the energy exchange with TKE transport equation. The rightmost  term on the right-hand side is the mean dissipation. Hence, $\frac{1}{2}\nu [\triangledown^2 {\widetilde w^2}]$ is the power input. By further manipulation, the power input for the whole system reads:

\begin{equation}\label{equ:power input}
{P_{in}} = 2\pi RL{\left. \nu{\left(\overline{\widetilde w\frac{{\partial \widetilde w}}{{\partial r}}}\right)} \right|_R}
\end{equation}

In consideration of constant body force, the power saving rate is defined as (\cite{gomez2016streamwise})
\begin{equation}\label{equ:power_saved}
S = ({P_c} - {P_u})/{P_u},
\end{equation}
in which $P$=$f_x$$\pi$$R^2$$LU_b$ is the power required to drive the flow and subscript $c$ and $u$ denotes controlled and uncontrolled pipe. A net energy saving rate $S$ is calculated by taking the power input into account:
\begin{equation}\label{equ:net_energy_saving}
N = ({P_c} - {P_u} - {P_{in}})/{P_u}.
\end{equation}
Finally, the effectiveness is computed as the ratio of power saved to the power input:
\begin{equation}\label{equ:effectiveness}
E = ({P_c} - {P_u})/{P_{in}}.
\end{equation}.

\begin{table}
\centering

\begin{tabular}{ccccc}
Case&$P_{in}$/$P_u$(\%)&$S$(\%)&$N$(\%)&$E$\\
\hline
1 & 67.22 & 31.47 & -35.75 & 0.47\\
2 & 32.86 & 207.7 & 174.8 & 6.32 \\
4 & 11.24 & 17.90 & 6.659 & 1.59\\
\end{tabular}
\caption{\label{tab:table2}Flow control energetic performance indices. Only one relaminarization case (case 2) is selected for assessment.}
\end{table}

The computed energetic performance indices are listed in table \ref{tab:table2}. Undoubtedly, in the constraint of CPG, relaminarization will produce a vast increase of mass flow rate (207.7\%) for case 2, thereby leading to a high net energy saving rate of 174.8\% since only 32.86\% of power input rate is required. This fact highlights that the ultimate target of flow control is reaching the laminar state (\cite{gatti2018global}). For non-relaminarization cases, case 1 achieves a larger power saving rate (31.47\%), but it is at the cost of high power input (67.22\%), which inevitably causes a negative net energy saving rate (-35.75\%). On the contrary, case 4 requires less power input (11.24\%) due to its low amplitude, and the power saving rate (17.9\%) is relatively low accordingly. But a positive net energy saving (6.659\%) could be achieved. Hence, in order to attain a positive net energy saving, it is recommended to lower the amplitude and increase the wavelength. In terms of effectiveness, case 4 (1.59) is three times as much as case 1 (0.47). The implication behind is that increasing the wavelength rather than increasing the amplitude seems to be a better choice to improve the control efficiency.

\section{Turbulent regime}\label{turbulent regime}

\subsection{ Basic flow statistics}\label{basic flow statistics}

\begin{figure}
\includegraphics[scale=0.6]{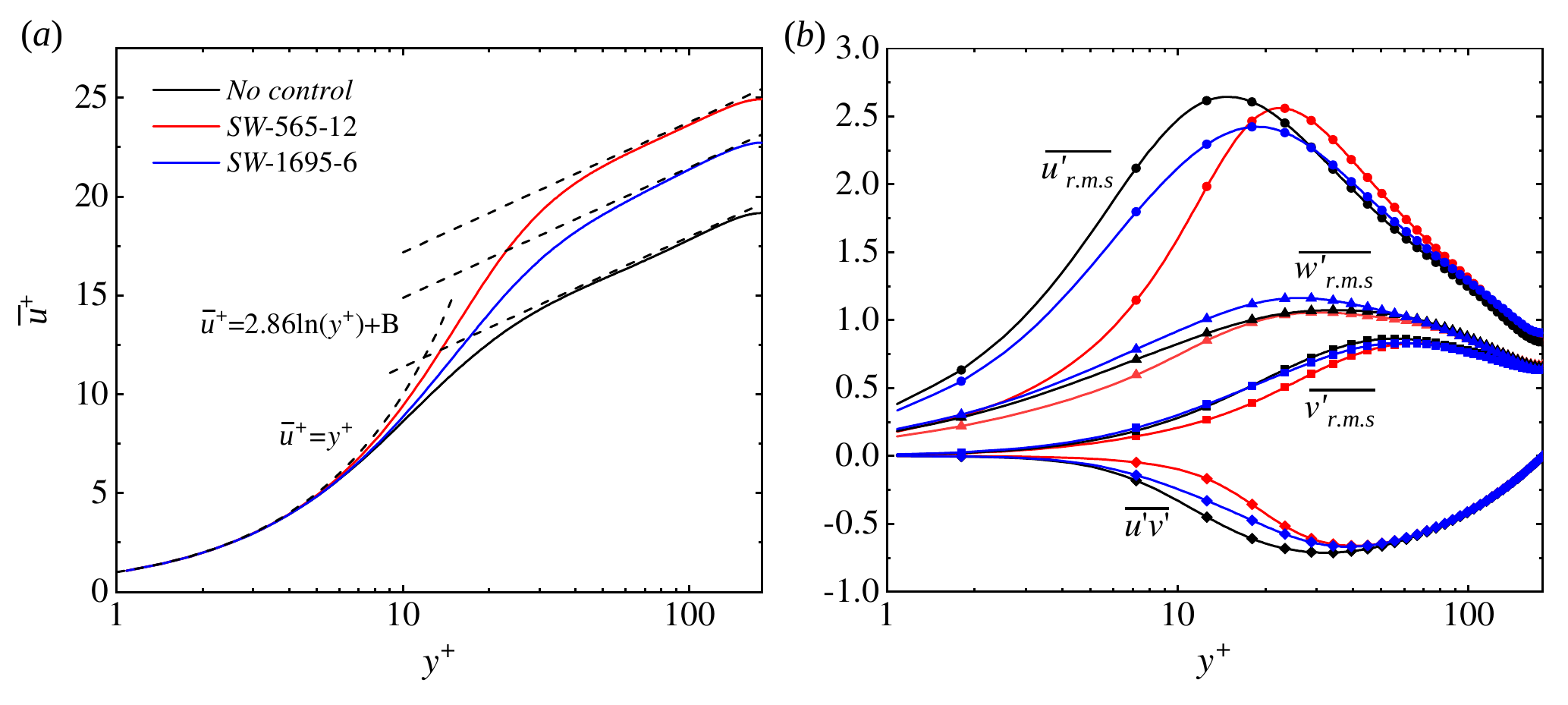}
\centering
\caption{\label{fig:mevel_ti} (\emph{a}) Wall-normal profiles of $\overline{u}$ for non-relminarization cases (case 1 and 4). The logarithmic law is adopted from \citet{eggels1994fully}. $B$=4.8 for uncontrolled case, $B$=8.3 for case 4 and $B$=10.6 for case 1. (\emph{b}) Wall-normal profiles of turbulence intensity and Reynolds shear stress for non-relaminarization cases. Black line: uncontrolled case; Red line: case 1; Blue line: case 4. Line with circles: $\overline{u'_{r.m.s}}$; line with triangles: $\overline{w'_{r.m.s}}$; line with squares: $\overline{v'_{r.m.s}}$; lines with diamonds: $\overline{u'v'}$. }
\end{figure}

This section presents the modification of basic flow statistics to illustrate the responses of fluids to standing wave wall oscillation. Profiles of streamwise mean velocity, scaled by the friction velocity $u_\tau$, are presented in figure \ref{fig:mevel_ti}(\emph{a}). Two types of dashed lines, which are the theoretical velocity distribution in viscous sublayer and logarithmic layer respectively, are included as reference. The logarithmic reference line is adopted from \citet{eggels1994fully}, in which the 'universal' constants for pipe flow is found to differ from that in channel. All profiles collapse in the viscous sublayer (0$\textless$$y^+$$\textless$5) due to the constraint of CPG. Parallel elevation can be observed in the outer layer with the increasing of drag reduction, indicating the increase of mass flow rate. Moreover, as the drag reduction increases, the logarithmic region shrinks accordingly, implying that the flow is driven towards the laminar state.

Profiles of turbulence intensity (i.e. Reynolds normal stress) as well as Reynolds shear stress are presented in figure \ref{fig:mevel_ti}(\emph{b}). When control is imposed, the general phenomenon, which has been reported broadly (\cite{skote2013comparison,duggleby2007effect,quadrio2000numerical}), is the decline of maximum streamwise stress with respect to the uncontrolled case, accompanied by a shift of wall-normal position of the maximum away from the wall. Interestingly, the maximum streamwise stress for case 1 is larger than case 4 despite that the drag reduction of case 1 is significantly higher than case 4, indicating that there is no direct correlation between the maximum streamwise stress and drag reduction. In the viscous sublayer, the streamwise stress reduces by at least 50\% for case 1 while the reduction for case 4 is comparatively small. For circumferential stress, there is a slight increase for case 4 across the buffer layer, where the near-wall streaks populates, while for case 1, on the contrary, it declines in the viscous sublayer and remains nearly unchanged beyond. Meanwhile, a close inspection of radial stress for case 4 shows that it increases slightly in the viscous sublayer and declines in the buffer layer, although the change is subtle. For case 1, the radial stress declines across the buffer layer. Unlike the irregular variation of Reynolds normal stress, the variation of shear stress clearly shows a monotonic trend, that is, it declines monotonically as the increasing of drag reduction. This is easily understood since shear stress appears explicitly in the integrated streamwise momentum equation. In the context of constant body force, the lower the shear stress, the larger the mass flow rate, hence the higher the drag reduction.

Based on the description above, two different drag-reduction scenario can be roughly depicted. For case 4, the thickness of SSL is large enough to affect the main part of near-wall streaks, causing the streaks to incline sinuously in streamwise direction, as will be shown later. The formation of sinuous pattern is then accompanied by an energy transfer from streamwise stress to circumferential stress (\cite{touber2012near}), resulting in the decline of streamwise stress and the elevation of circumferential stress. Possibly due to the low amplitude, the streak inclination is not intense enough to induce a remarkable change in radial stress, whose physical interpretation is the near-wall downward or upward turbulent motions. Nevertheless, the sinuous streak pattern indeed lower the magnitude of shear stress and achieve drag reduction. For case 1, the thickness of SSL is too small so that only a small fraction of the near-wall streaks can be affected. In this way, no extensive streak inclination occurs and the SSL could be regarded as a shearing layer that lies between the wall and the flow structures, preventing the near-wall splatting or anti-splatting, which is an important source of the shear stress. Therefore, the shear stress declines. Within the SSL, the turbulence intensity is attenuated dramatically due to the strong shearing, and hence all Reynolds stress declines accordingly.  

\begin{figure}
\includegraphics[scale=0.6]{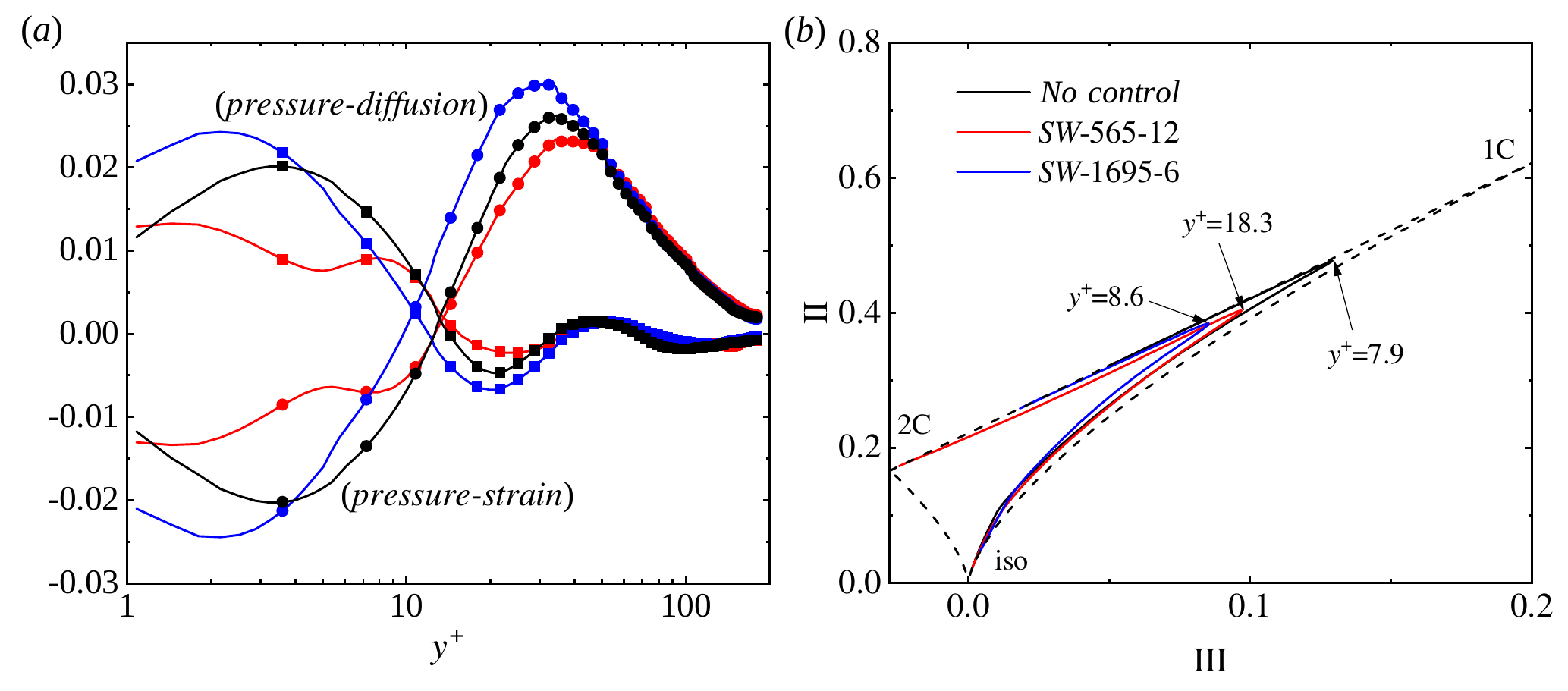}
\centering
\caption{\label{fig:ani_rapre} (\emph{a}) Wall-normal profiles of pressure-diffusion and pressure-strain term of radial stress $\overline{v'v'}$. Black lines: uncontrolled case; Blue lines: case 4; Red lines: case 1. Lines with squares: pressure-diffusion term; Lines with circles: pressure-strain term. (\emph{b}) Anisotropy-invariant maps for different cases. $a_{ij}$=$\overline{u'_i}$$\overline{u'_j}$/$\overline{u'_i}$$\overline{u'_i}$-$\delta_{ij}$/3, II=$a_{ij}$$a_{ji}$,III=$a_{ij}$$a_{jk}$$a_{ki}$, where the conventional tensor summation is employed and $i$,$j$=1,2,3 correspond to $r$,$\theta$,$z$ direction respectively. Dashed curve lines: II=$\frac{3}{2}$($\frac{4}{3}$$\vert$III$\vert$)$^\frac{2}{3}$; Dashed straight line: II=$\frac{2}{9}$+2III. 1C and 2C denotes one- and two-component turbulence respectively and iso denotes isotropic turbulence.   }
\end{figure}

In order to verify the scenario depicted above, the pressure-velocity gradient term in the transport equation of $\overline{v'v'}$ is examined, as shown in figure \ref{fig:ani_rapre}(\emph{a}). It is known that the pressure-velocity gradient can be split into pressure-diffusion and pressure-strain fragments, among which the role of the latter is to drain energy from $\overline{v'v'}$ and redistribute to the rest two stress components through near-wall splatting effect (\cite{eggels1994fully,mansour1988reynolds}). These two terms are of the opposite sign and virtually cancel each other near the wall. As can be seen, compared to uncontrolled case, the magnitude of these two terms are significantly attenuated within the viscous sublayer for case 1, especially around $y^+$=5, where a local minimum can be observed. Such local minimum clearly implies a resisting effect of the penetration of outer eddies into the near-wall layer. While on the contrary, the magnitude is enhanced for case 4 across the viscous sublayer, indicating a more enhanced wall-normal mixing in the immediate wall region. 

The modification of near-wall turbulent state can be further examined with the aid of anisotropy map since the streaky structure is inherently anisotropic, as conveyed in figure \ref{fig:ani_rapre}(\emph{b}). Details of drawing the figure can be referred to \citet{frohnapfel2007interpretation}. As can be seen, once the drag reduction is achieved, the inflection point moves away from the one-component turbulence, indicating an attenuation of the strength of near-wall streaks. The fact that the distance from the inflection point to one-component turbulence for case 4 is farther than case 1 implies that the streaky structures of case 1 is less affected by control. Moreover, the wall-normal distance of inflection point to the wall for case 1 is significantly larger than case 4 and uncontrolled case, while for the latter two cases, it is nearly identical, indicating that the near-wall streaks are elevated by the SSL for case 1 and hence providing further evidence for the drag-reduction scenario described above. In the vicinity of the wall, the turbulence state for two cases both moves towards to two-component turbulence, with case 1 being more intense, which, again, can be attributed to the more strong shearing nature of SSL.   

\subsection{Spatial Stokes layer}\label{spatial stokes layer}

It is known that there exists an exact solution when static fluids are bounded by temporal oscillating flat plate, which is the so-called Stokes second problem. Similarly, based on an assumption that the thickness of spatially oscillating boundary layer is much small compared with the channel width, \cite{viotti2009streamwise} derived an analytical solution for laminar channel flow with spatial wall oscillation imposed, whose accuracy is demonstrated by \cite{skote2013comparison} for flat turbulent boundary layer. For boundary condition of equation (\ref{equ:bc}), the laminar solution reads:
\begin{equation}\label{equ:laso}
\widetilde {w}(x,y) = \frac{{{A}}}{{Ai(0)}}{\mathop{\Re}\nolimits} [e^{ikx}Ai( - i\frac{y}{{{\delta _x}}}{e^{ - \frac{{4\pi }}{3}i}})],
\end{equation}
where $Ai$ is the Airy function and the characteristic thickness of boundary layer can be written as:
\begin{equation}\label{equ:thickness}
{\delta _x} = {(\frac{{{\lambda}{\nu ^2}}}{{2\pi u_\tau ^2}})^{\frac{1}{3}}}.
\end{equation}

\begin{figure}
\centering
\includegraphics[scale=0.65]{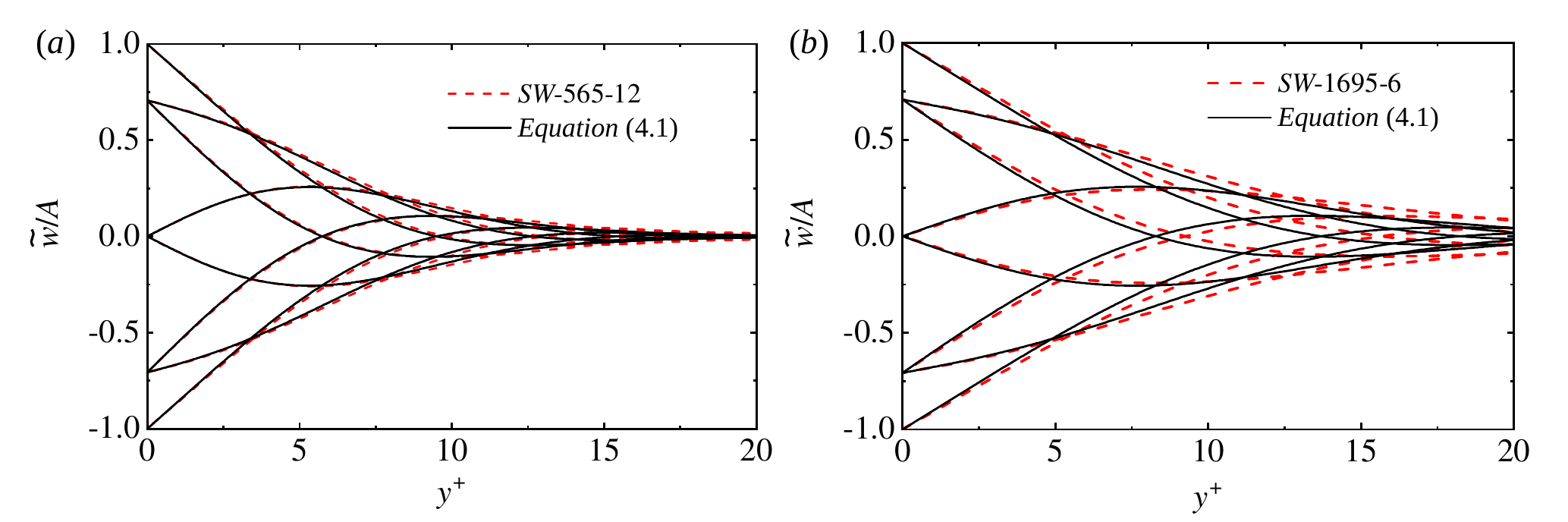}
\caption{\label{fig:spatial stokes layer} The spatial Stokes layer (SSL) and wall-normal profiles of $\widetilde{w}$ at different phases. (\emph{a}) case 1. (\emph{b}) case 4.}
\end{figure}

\begin{figure}
\centering
\includegraphics[scale=0.65]{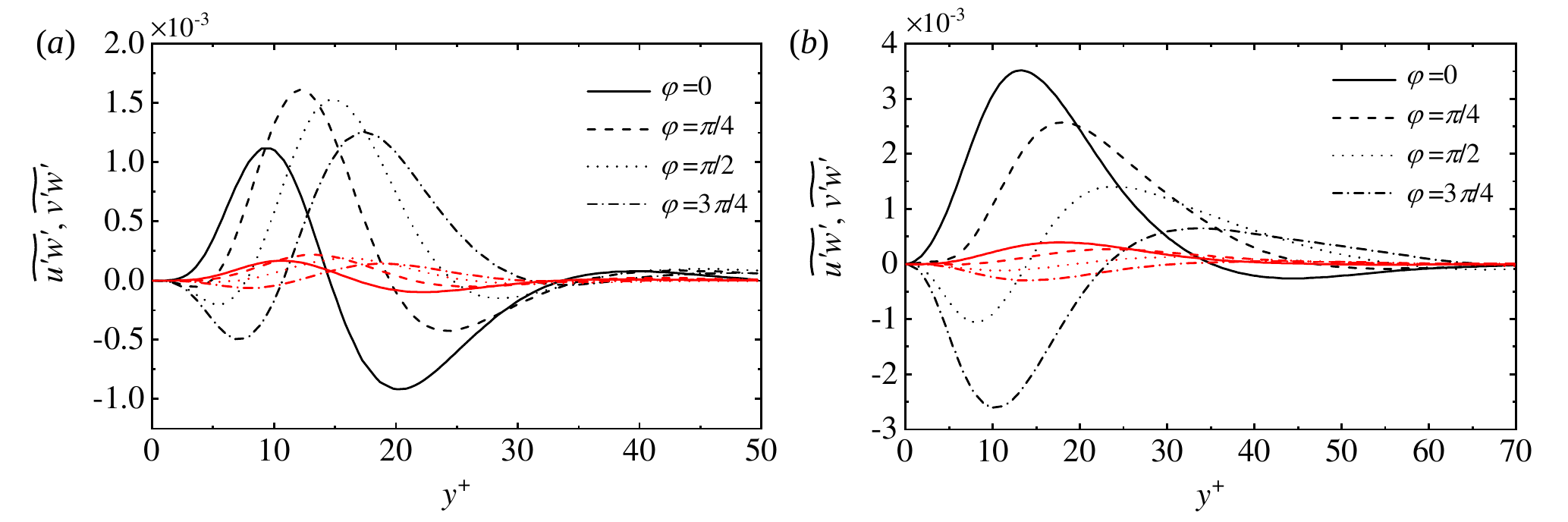}
\caption{\label{fig:spatial stokes layer2} Wall-normal profiles of Reynolds stress $\widetilde{u'w'}$ and $\widetilde{v'w'}$ at different phases $\varphi$. Only half wavelength is shown for brevity. Black lines: $\widetilde{u'w'}$, Red lines: $\widetilde{v'w'}$ (\emph{a}) case 1. (\emph{b}) case 4.}
\end{figure}

In terms of cylindrical geometry, scenario can be significantly different due to the curvature effect, which manifests itself as the coupling of circumferential and radial velocity. For temporal wall oscillation control, it is shown that for Reynolds number close to present study, a Stokes layer thickness of $\delta^+$$\approx$35 will induce small curvature effect, leading to excellent agreement between the laminar solution for flat plate and turbulence results of circular pipe (\cite{quadrio2000numerical}). Thus, it is interesting to evaluate the degree of closeness between present DNS results of turbulent flow and laminar analytical solution of equation (\ref{equ:laso}).

The governing equation of $\widetilde{w}$ for circular pipe follows from:

\begin{equation}\label{equ:spanwise momentum}
\underbrace{\widetilde {v}\frac{{\partial \widetilde {w}}}{{\partial r}} + \frac{{\widetilde {v}\widetilde {w}}}{r}}_{\textsc{1}} + \underbrace{\widetilde {u}\frac{{\partial \widetilde {w}}}{{\partial x}}}_{\textsc{2}} + \underbrace{\frac{{\partial \widetilde {v^{'}w ^{'}}}}{{\partial r}} + \frac{{2\widetilde {v^{'}w ^{'}}}}{r}}_{\textsc{3}} + \underbrace{\frac{{\partial \widetilde {u^{'}w ^{'}}}}{{\partial x}}}_{\textsc{4}} \\
= \underbrace{\nu (\frac{{{\partial ^2}\widetilde {{w }}}}{{\partial {r^2}}} + \frac{{{\partial ^2}\widetilde {w}}}{{\partial {x^2}}} + \frac{1}{r}\frac{{\partial \widetilde {w}}}{{\partial r}} - \frac{{\widetilde {w}}}{{{r^2}}})}_{\textsc{5}}.
\end{equation}

Term 1 is associated with the curvature effect since $\widetilde{v}$ is zero for flat plate while term 2 denotes the mean convection in streamwise direction. Sum of terms 3 and 4 represents the turbulence modulation and term 5 is the viscous term. Terms 1,2 and 5 lead to the laminar solution for circular pipe, which will turn into equation (\ref{equ:laso}) if the curvature effect is small enough. Thus, the deviation of present DNS results from equation (\ref{equ:laso}) is rooted in the curvature effect and turbulence modulation for which it can be qualitatively estimated with the aid of following dimensional analysis.

According to figure \ref{fig:spatial stokes layer}, the thickness of SSL is approximately less than 35, hence the curvature effect can be neglected and term 1 is not considered (In fact, we have examined the magnitude of term 1, which is proved to be orders of magnitude smaller than other terms). The characteristic length in streamwise direction can be taken as ${\lambda}$ since the streamwise periodicity of control law (\ref{equ:bc}) and the velocity scale of $\widetilde{w}$ is $A$. As shown in \cite{viotti2009streamwise}, thickness equation of (\ref{equ:thickness}) provides an universal representation of scaling properties for turbulent flow. Hence, $\delta_x$ is applicable for the wall-normal length scale. The mean streamwise velocity grows linearly along the wall-normal direction in the viscous sublayer, thus $\widetilde{u}$ can be considered as the same order of magnitude within the SSL for all cases. $\partial ^2 \widetilde{w}/\partial x^2$ is comparatively negligible because ${\lambda}$ is much larger than $\delta_x$. As a consequence, we obtain:
\begin{equation}\label{equ:order of magnitude}
2 \sim \frac{{\overline{u}A}}{{{\lambda}}},3 \sim \frac{{{\varepsilon _1}}}{{{\delta _x}}},4 \sim \frac{{{\varepsilon _2}}}{{{\lambda}}},5 \sim \frac{{\nu {A}}}{{\delta _x^2}}
\end{equation}
where $\varepsilon_1$ and $\varepsilon_2$ characterize the magnitude of $\widetilde {v^{'}w ^{'}}$ and $\widetilde {u^{'}w ^{'}}$ respectively. Therefore, term 2 is our focus. Obviously, high amplitude and low wavelength tend to increase the magnitude of term 2, to which the turbulence modulation can be relatively small. The magnitude of Reynolds stress for cases 1 and 4, shown in figure \ref{fig:spatial stokes layer2}, results in a turbulence modulation of order of magnitude of $10^{-2}$ for both two cases. As for term 2, DNS data gives order of magnitude of $10^{-1}$ for case 1 and $10^{-2}$ for case 4. Hence, large wavelength in conjunction with small amplitude tends to lower the effect of mean convection, thereby the impact of turbulence develops relatively, which finally causes the distinct deviation. This is illustrated by figure \ref{fig:spatial stokes layer}(\emph{a})(\emph{b}), in which good overlap between equation (\ref{equ:laso}) and present DNS data is found for case 1 while noticeable departure can be observed for case 4. Note that there is significant difference of the wall-normal distribution of phase-averaged Reynolds stress components between low- and large-wavelength control (figure \ref{fig:spatial stokes layer2}). This is due to the streamwise mean convection, which enhances the impact of upstream phase on downstream phase, especially for the low-wavelength control. Moreover, large wavelength tends to induce larger value of Reynolds stress $\widetilde{u^{'}w^{'}}$ and $\widetilde{v^{'}w^{'}}$, among which $\widetilde{u^{'}w^{'}}$ is significantly larger than $\widetilde{v^{'}w^{'}}$. However, the fact that $\lambda\gg\delta_x$ significantly diminishes the magnitude of term 4, making it acts as the same order of magnitude as term 3 or even much smaller according to (\ref{equ:order of magnitude}).

\subsection{One-dimensional spectra}\label{spectra}

\begin{figure}
\centering
\includegraphics[scale=0.65]{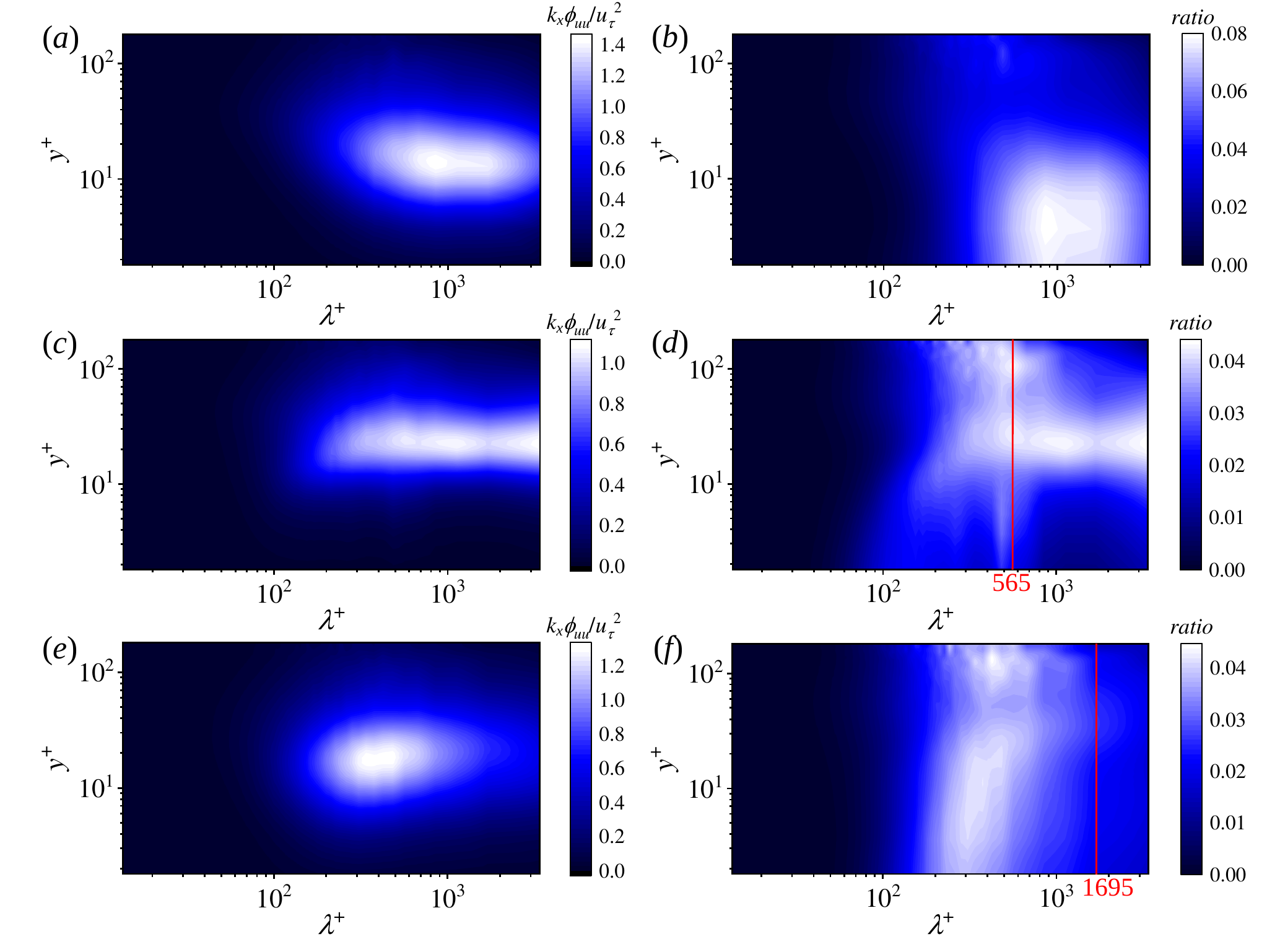}
\caption{\label{fig:spectra} One-dimensional premultiplied spectra maps of streamwise fluctuation, $k_x \phi_{uu}$. Note that $\overline {u'u'}  = \int_0^\infty  {{\phi _{uu}}d{k_x}}$ and $k_x$=2$\pi$/$\lambda$. The left column gives the absolute value of $k_x \phi_{uu}$ nondimensionalized by $u_\tau$. The right column shows the relative value, which is calculated by absolute value divided by
the total resolved energy at corresponding wall-normal position. For controlled cases, the corresponding control wavelength is marked by vertical red line. (\emph{a})(\emph{b}) uncontrolled case. (\emph{c})(\emph{d}) case 1. (\emph{e})(\emph{f}) case 4.}
\end{figure}

Contours of streamwise premultiplied one-dimensional spectra as a function of wall-normal position and wavelength are shown in figure \ref{fig:spectra}. This sort of spectra map is generally employed to identify the large and very-large scale motion at high Reynolds number (\cite{wu_baltzer_adrian_2012,guala2006large}). Rather, large-scale motions is not our focus since the Reynolds number considered here is relatively low. In consideration of the spatial periodicity of standing wave, it is interesting to examine the relationship between control wavelength and near-wall dominant flow structures. There are two kinds of contour map for each case in figure \ref{fig:spectra}, with the left column giving the absolute value and the right presenting the value normalized by total resolved energy at corresponding wall-normal position. In this way we can not only identify the most energetic region in total flow field, but also assess the relative importance of different scales of motions at different wall-normal positions. The respective control wavelength are marked by red vertical line.

For uncontrolled pipe, the most energetic wavelength is of order of $O$($10^3$) wall units, which coincides with the commonly accepted length scale of flow structures at $y^+$$\approx$15. From the perspective of relative value, uniform peak of wavelength of $O$($10^3$) occurs below $y^+$$\approx$15, highlighting the dominant role of such scale of motions. As the wall-normal position moves beyond $y^+$$\approx$15, the distinct predominance of that wavelength gradually weakens, indicating a more isotropic turbulence state. 

With control imposed, there is a clear broadening of absolute energy peak, whose wall-normal location moves outwards to $y^+$$\approx$20, in wavelength for case 1, as shown in figure \ref{fig:spectra}(\emph{c}). Note that $y^+$$\approx$20 is well beyond the SSL of case 1 (figure \ref{fig:spatial stokes layer}(\emph{a})), hence such broadening effect could be attributed to the increase of mean velocity gradient in this region, as shown in figure \ref{fig:mevel_ti}(\emph{a}). The significantly enhanced mean velocity gradient stretches the near-wall structures in streamwise direction, after which partial energy is allocated to the larger scale of motions. Interestingly, below $y^+$$\approx$15, one dominant wavelength, which peaks in relative value, is very close to the control wavelength (red line), highlighting the close correspondence between the most energetic scale of motion (in relative sense) and control wavelength within the SSL. For case 4, elevation of wall-normal position of absolute energy peak is attenuated compared to case 1, and the most energetic wavelength is reduced to approximately $\lambda^+$=400. Similarly, the relative energy also peaks at around $\lambda^+$=300-400 within the SSL, which is far less than the control wavelength $\lambda^+$=1695. In the central region, the energy distribution is of great resemblance for controlled and uncontrolled cases, for which the most energetic motion (in relative sense) locates at $\lambda^+$=400-500, indicating that the far outer layer is unaffected by the wall motion. Thus, it can be summarized that for the low wavelength control (case 1), the dominant scale of motions is closely related to the control wavelength within the SSL, while for the large wavelength control (case 4), such correlation is rather weak, with the dominant scale of flow motions being far less than the control wavelength.

\subsection{\label{flow structures} Flow structures}

It is well-established that in wall-bounded turbulence, the near-wall region is inundated with alternately low- and high-speed region, which can be interpreted as ’streak’, accompanied with streamwise vortices that are in close relation to the ejection and sweep events and hence the self-sustaining process. It is unambiguous that the wall motion will destroy the original correlation and modify the near-wall flow structures. In this section, we present the visualization of near-wall low-speed streaks to exhibit the flow state under the effect of standing wave wall motion. 

\begin{figure}
\centering
\includegraphics[scale=0.65]{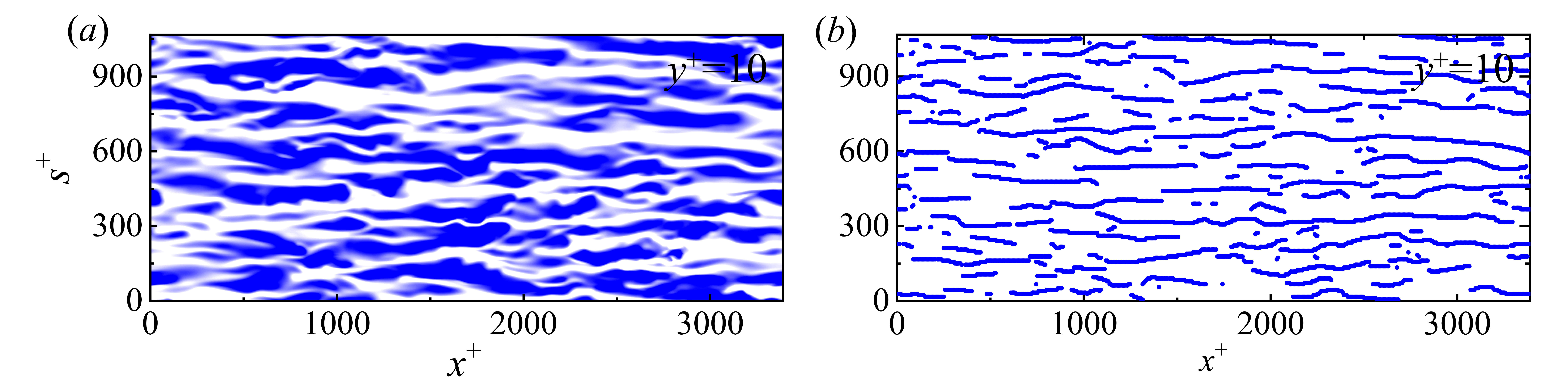}
\caption{\label{fig:nocontrol fs} Instantaneous field of streamwise velocity fluctuation $u'$ of uncontrolled case at wall-normal position of $y^+$=10. The snapshot is constructed by unfolding the cylinder plane and $s$=$r\theta$ represents the arc length. The flow direction is from left to right. (\emph{a}) Contour map of $u'$, in which the data are clipped to $u'$$\in$[-0.1,0.1]. White regions correspond to the velocity deficit while blue regions denote the excess. (\emph{b}) ’Spine’ plot of the same field in (\emph{a}).  }
\end{figure}

Typical pattern of near-wall low-speed streaks at $y^+$=10 is shown in figure \ref{fig:nocontrol fs}(\emph{a}) for uncontrolled pipe. Such a picture, in which the contour of streamwise velocity fluctuation is presented, is constructed by unfolding the $x$-$\theta$ plane at specified wall-normal location. Several low-speed streaks, with an average separating distance of $O$($10^2$), meander downstream. In order to provide a more clear representation of near-wall streaks, the ’spine’ of low-speed streaks are plotted in figure \ref{fig:nocontrol fs}(\emph{b}). The ’spine’ is extracted by taking the location of local minimum of negative streamwise velocity fluctuation that enclosed in low-speed region at many $x$ locations (\cite{dennis2011experimental2}), and consequently only the information of streamwise length is preserved. It can be found that the very long streak observed in figure \ref{fig:nocontrol fs}(\emph{a}) is actually the alignment of several shorter individual streaks and the flow field is occupied mostly by streaks whose length is of order of $O$($10^3$) wall units, consistent with the results of one-dimensional spectra in figure \ref{fig:spectra}(\emph{a}), with some fragmental streaks alongside.

\begin{figure}
\centering
\includegraphics[scale=0.65]{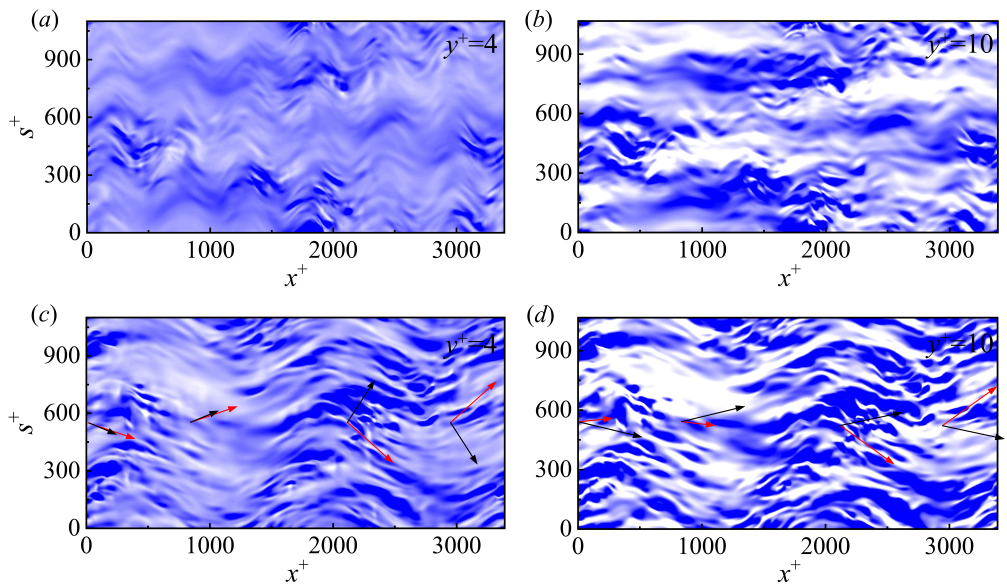}
\caption{\label{fig:instantaneous field} Instantaneous snapshots of streamwise velocity fluctuation $u'$ at wall-normal position of $y^+$=[4,10]. The snapshot is constructed by unfolding the cylinder plane and $s$=$r\theta$ represents the arc length. The data of contour map are clipped to $u'$$\in$[-0.1,0.1]. The flow direction is from left to right. White regions correspond to velocity deficit while blue regions represent the excess. (\emph{a})(\emph{b}) case 1. (\emph{c})(\emph{d}) case 4. For (\emph{c})(\emph{d}), additional vectors at four phases, which, from the left to the right, are $x$=$0,\lambda/2,\lambda/4,3\lambda/4$, are shown. Red arrows represent the local shear force vectors and black arrows correspond to the local mean velocity vectors.}
\end{figure}

\begin{figure}
\centering
\includegraphics[scale=0.65]{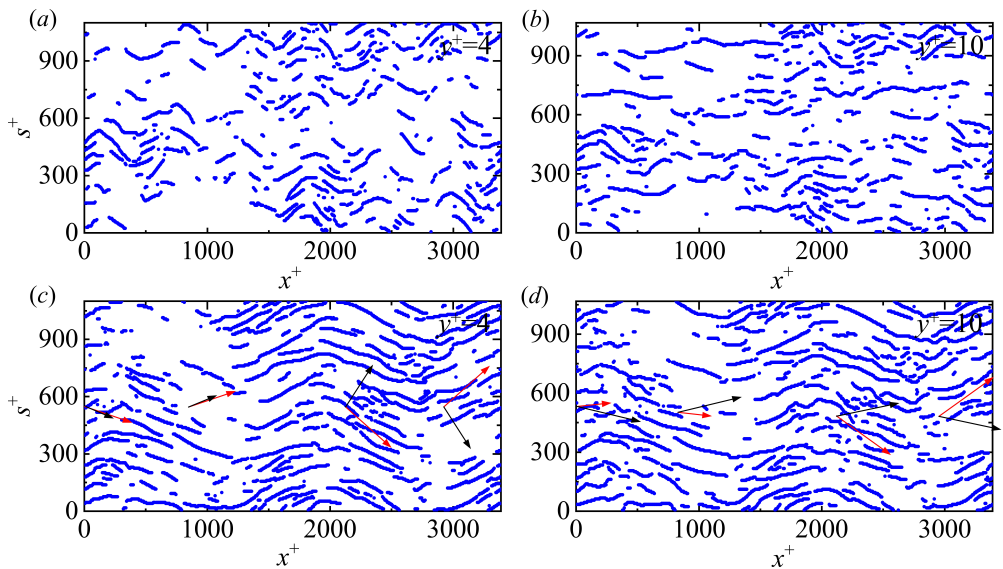}
\caption{\label{fig:spine} 'Spine' plots of the same instantaneous field in figure \ref{fig:instantaneous field}. (\emph{a})(\emph{b})case 1. (\emph{c})(\emph{d})case 4.  }
\end{figure}

When control is imposed, the streamwise wavy pattern of streaks can be clearly observed for case 1 at $y^+$=4 (figure \ref{fig:instantaneous field}(\emph{a})), with the appearance being blurred. As the wall-normal location moves to $y^+$=10, the wavy pattern is less distinct and the separating distance enlarges (figure \ref{fig:instantaneous field}(\emph{b})). Figure \ref{fig:spine}(\emph{a})(\emph{b}) shows the ’spine’ plot version of the same flow field. The wavy pattern at $y^+$=4 can be clearly reproduced. At $y^+$=10, the 'spine' plot shows that most of streaks recover its straight orientation but possess a shorter length compared with uncontrolled case, indicating that the original long straight low-speed streaks break into several pieces, whose length is closely related to the control wavelength. As for case 4, similar wavy pattern can also be found at both $y^+$=4 and 10 but with a larger streamwise scale. The larger control wavelength produces larger thickness of SSL, hence expanding the wall-normal scope of wavy streaks. As can be seen from the ’spine’ plot, such ’breaking’ effect is relatively weak as there are many long meandering inclined streaks, whose length is examined to be shorter than that in uncontrolled case as evident in figure \ref{fig:spectra}(\emph{e}), in which the near-wall peak is located at a smaller wavelength.

A purely visual inspection of figure \ref{fig:instantaneous field},\ref{fig:spine} seems to suggest that the streamwise wavy streaks conform to the wall velocity since the streak inclination is visually consistent with the graph of wall velocity shown in figure \ref{fig:computational domain}. But in fact, the local mean circumferential velocity is a function of both the wall-normal and streamwise location. Moreover, the generated strong shear forces would also play an important role. Hence, following \cite{touber2012near}, the correspondence between the local mean shear force, local mean velocity and streak orientation is explored with the aid of the same snapshots for case 4. The choice of this case is motivated by the fact that the distinct near-wall streaky structures under this condition allows us to make unambiguous comparison. The directions of local shear forces and mean velocity at four phases, which in sequence are $x$=0,$\lambda$/4,$\lambda$/2,3$\lambda$/4, covering a whole wavelength, are shown by the arrows. Black arrows represent the local velocity vector [$u$,$w$] while red arrows correspond to the local mean shear force vector [$\partial \widetilde{u}$/$\partial r$,$\partial \widetilde{w}$/$\partial r$+$\partial \widetilde{w}$/$\partial x$]. In fact, the two components of mean shear force that are associated with $\widetilde{w}$ could alternatively possess the same direction or the opposite, but $\partial \widetilde{w}$/$\partial x$ is usually one order of magnitude lower than $\partial \widetilde{w}$/$\partial r$ due to the large difference in scale of SSL in streamwise and wall-normal direction. Hence these two components are merged as a resultant force. Again, ’spine’ plot allows for a more clear inspection.

At $y^+$=4, the shear force and velocity vector both point to a direction that forms an angle to the streak orientation when the wall velocity is zero ($x$=0,$\lambda$/2). On the contrary, when the wall velocity reaches its maximum ($x$=$\lambda$/4,3$\lambda$/4), their direction forms a obtuse angle, among which the shear vector seems to be aligned with the streak orientation. Note that these vectors are representing the directions at a single streamwise location, hence it can be interpreted that the shear force twists the streak by an angle. This scenario also suits for $y^+$=10, but the difference is that the shear vector changes direction and is aligned with the streak orientation when the wall velocity is zero while the mean velocity does not. Such change of direction of shear results from the non-monotonic wall-normal variation of $\widetilde{w}$ at these phases, as shown in figure \ref{fig:spatial stokes layer}(\emph{b}). Overall, the shear force beyond viscous sublayer dominates the streak orientation and the streak orientation lags behind the wall motion in streamwise direction. It is tempting to speculate that such sinuously inclined pattern disrupts the downstream development of near-wall flow structures, which is reflected by the shortening of streamwise scale, thereby leading to the decline of shear stress.

\section{Relaminarization}\label{relaminarization}

In this section, we focus on the relaminarization cases (case 2 and 5) to investigate the transient dynamics during the relaminarization process, in an effort to explore the mechanism behind. Specifically, we aim to figure out what causes the relaminarization in turbulent pipe flow as no relaminarization occurs in channel at the same control parameters and at similar Reynolds number (\citet{quadrio2009streamwise,viotti2009streamwise}). To this end, we additionally performed a DNS of turbulent channel flow at $Re_\tau$=180 with the control parameter as same as case 2. Through the comparison between channel and pipe, we were able to gain some insights into the relaminarization mechanism.

\subsection{Transient dynamics in relaminarization process}\label{flow dynamics}

\begin{figure}
\centering
\includegraphics[scale=0.65]{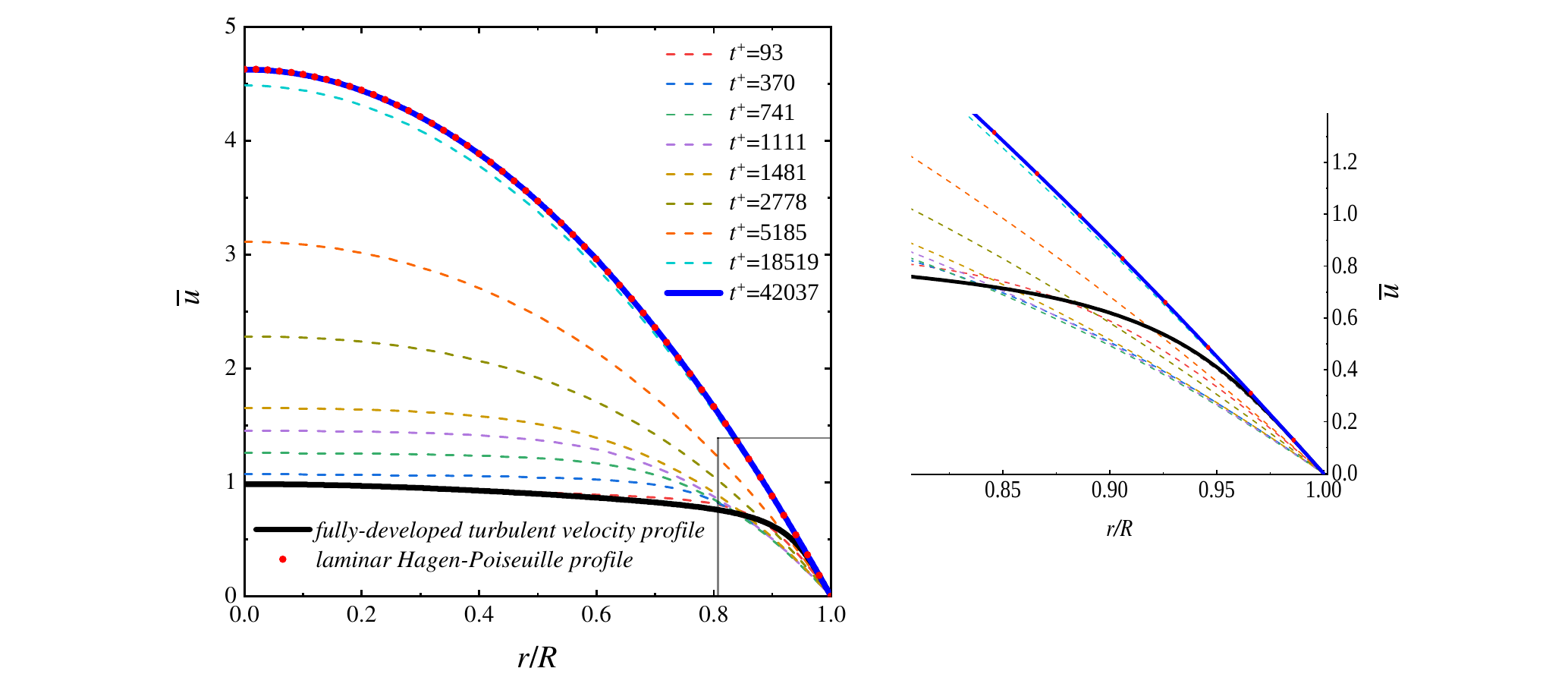}
\caption{\label{fig:meanvelocity evolution} Temporal evolution of $\overline{u}$ for case 5. Enlarged view of grey box is presented on the right.}
\end{figure}

\begin{figure}
\centering
\includegraphics[scale=0.65]{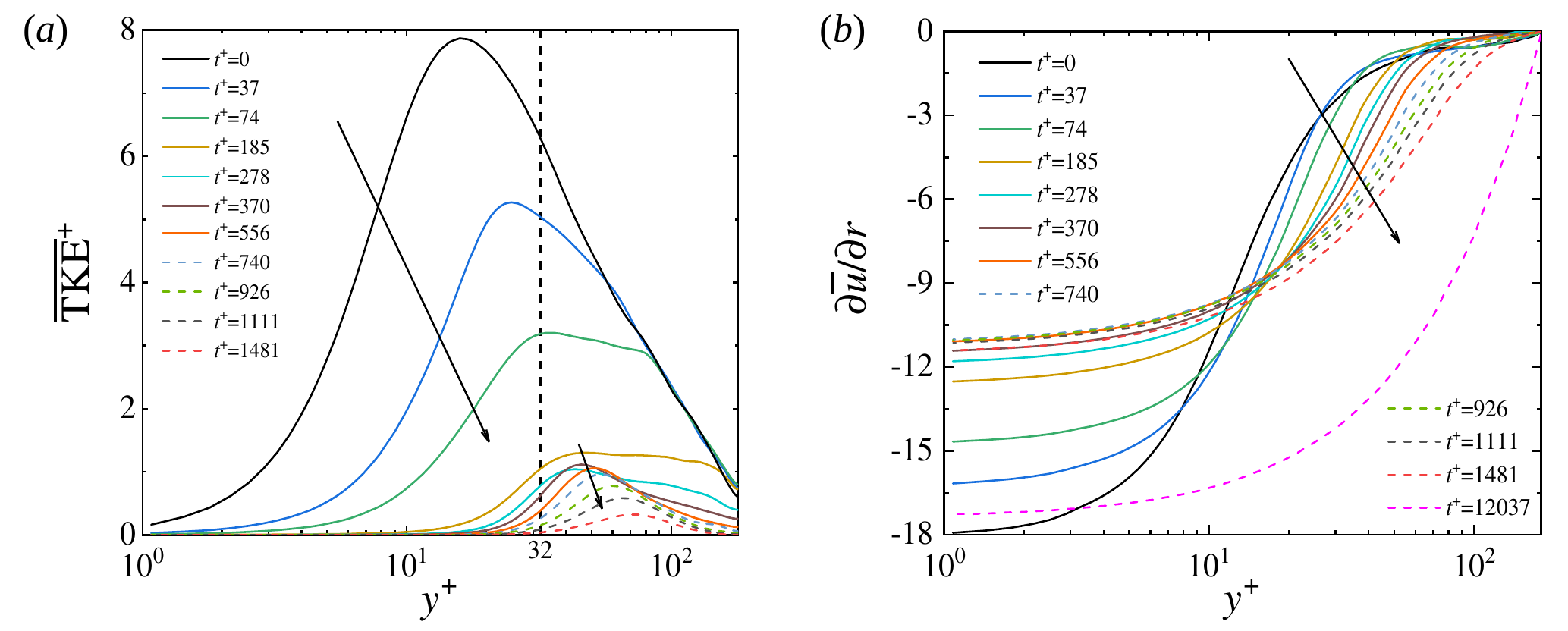}
\caption{\label{fig:TKE_grad} (\emph{a}) Temporal evolution of TKE. (\emph{b}) Time evolution of mean velocity gradient $\partial \overline{u}$/$\partial r$. Black vertical dashed line denotes the thickness of SSL at $t^+$=741, which is the largest value during the relaminarization. The arrows denote the direction of change in time.}
\end{figure}

\begin{figure}
\centering
\includegraphics[scale=0.65]{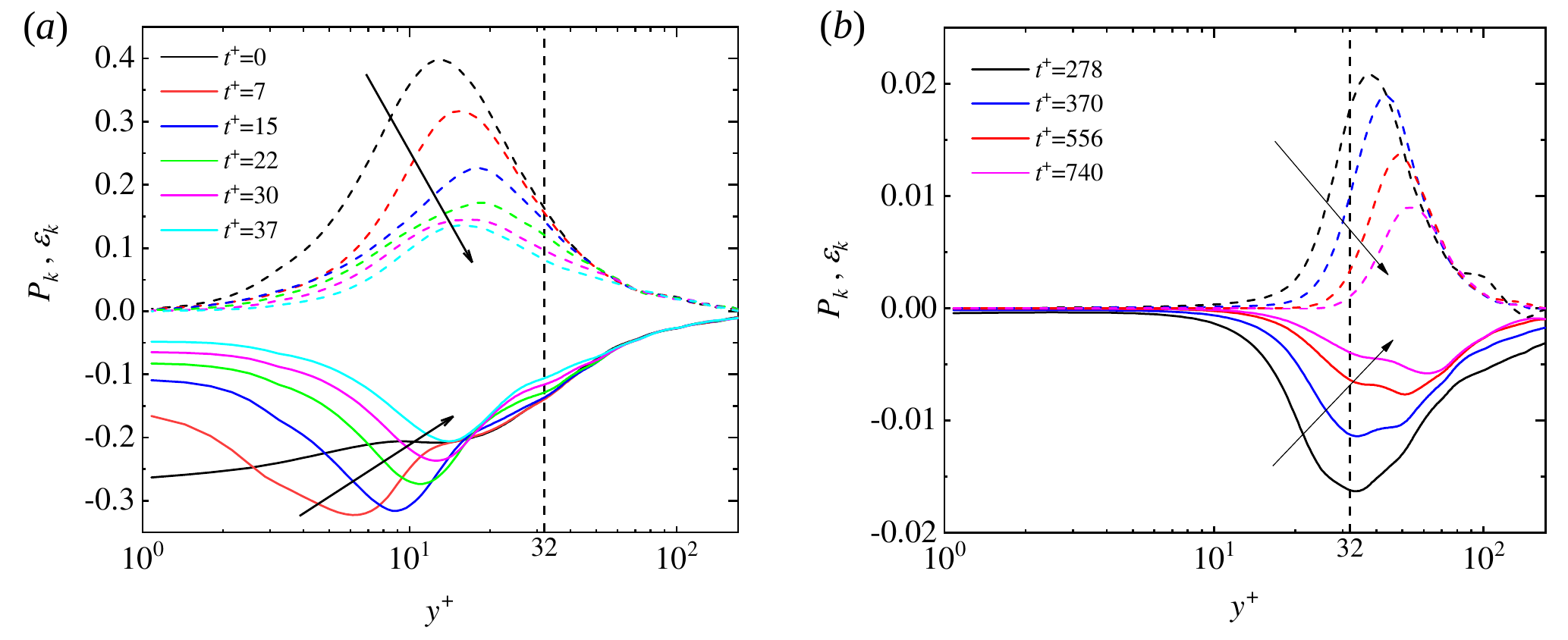}
\caption{\label{fig:TKE_pd} Temporal evolution of production (dashed lines) and dissipation (solid lines) of TKE budget. The arrows denote the direction of change in time. As depicted in the text, (\emph{a}) corresponds to the first stage while (\emph{b}) is associated with the second stage.}
\end{figure}

\begin{figure}
\centering
\includegraphics[scale=0.65]{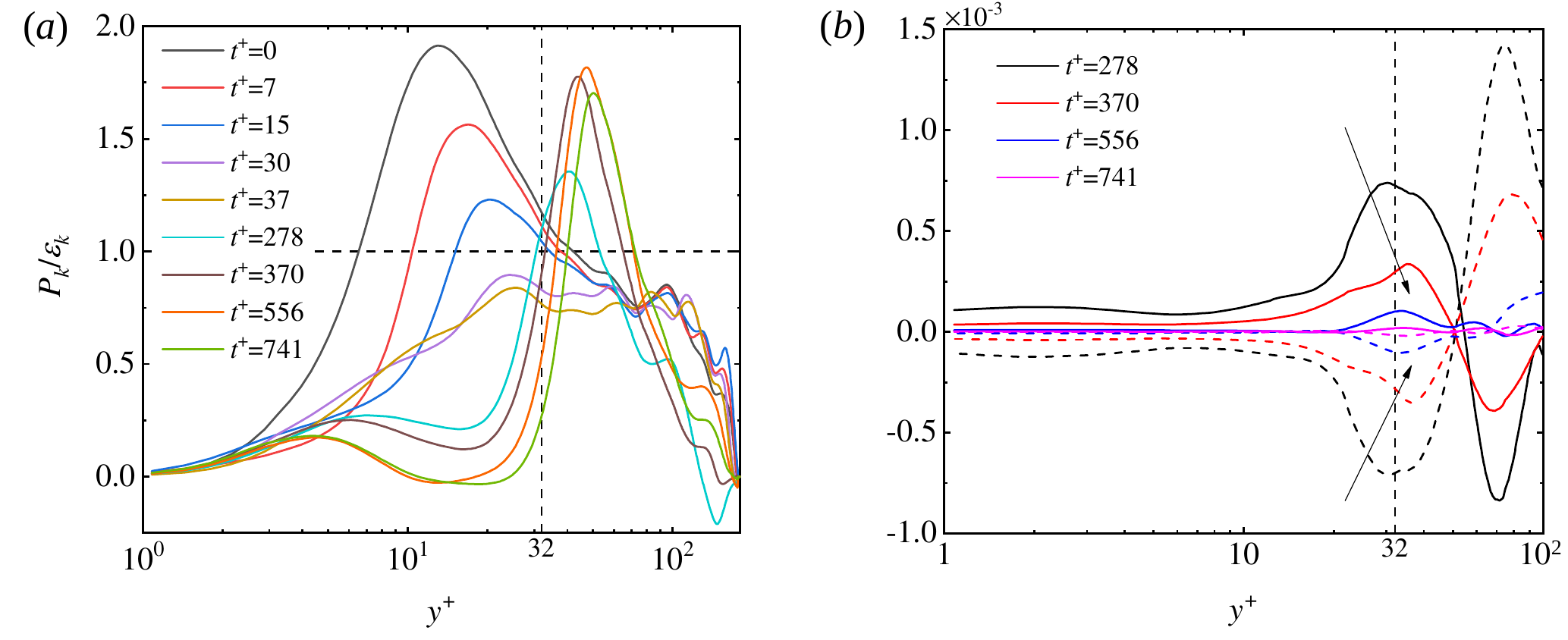}
\caption{\label{fig:ratio_pre} (\emph{a}) Time evolution of the ratio of production to dissipation ($P_k$/$\epsilon_k$). The arrows denote the direction of change in time. (\emph{b}) Time evolution of pressure diffusion (solid lines) and pressure strain term (dashed lines) of $\overline{v'v'}$ budget. Black vertical dashed line denotes the thickness of SSL at $t^+$=741, which is the largest value during the relaminarization.}
\end{figure}

In this subsection, we focus on the transient dynamics in the relaminarization process of case 5. Figure \ref{fig:meanvelocity evolution} presents the time evolution of streamwise mean velocity profile $\overline{u}$($r$). Control is imposed at $t^+$=0. As expected, the profile evolves from an initial flat shape (black solid line) to the final parabolic shape (blue solid line), which matches well with the analytical laminar Hagen-Poiseuille profile in pipe, indicating a fully relaminarization. Due to the CPG constraint, the fully-developed turbulent profile collapses with the laminar Hagen-Poiseuille profile in the vicinity of the wall (0.96$\textless$$r$/$R$$\textless$1) (\citet{marusic2007laminar}). The mean velocity in the core region increases monotonically while a careful examination of near-wall region reveals a non-monotonic trend (see the enlarged view of the grey box in figure \ref{fig:meanvelocity evolution}). In the region of (0.85$\textless$$r$/$R$$\textless$1), $\overline{u}$ starts to decrease after control is imposed ($t^+$=0) and reaches minimum at approximately $t^+$=741, after which it rises slowly until the final laminar Hagen-Poiseuille profile is reached. This leads to the non-monotonic change of velocity gradient at the wall, as shown in figure \ref{fig:TKE_grad}(\emph{b}), that is, it decreases initially and then increases to the final steady value. The implication behind is that the thickness of SSL increases initially and then decreases according to equation (\ref{equ:thickness}). Hence, in figure \ref{fig:TKE_grad}(\emph{a}), \ref{fig:TKE_pd} and \ref{fig:ratio_pre}, we mark the largest SSL thickness ($y^+$=32) by black vertical dashed line. The largest thickness is calculated by finding the wall-normal location, from the wall towards the center, where $\vert$$\widetilde{w}$$\vert$ decreases below 1\% of the velocity amplitude at $t^+$=741.

\begin{figure}
\centering
\includegraphics[scale=0.67]{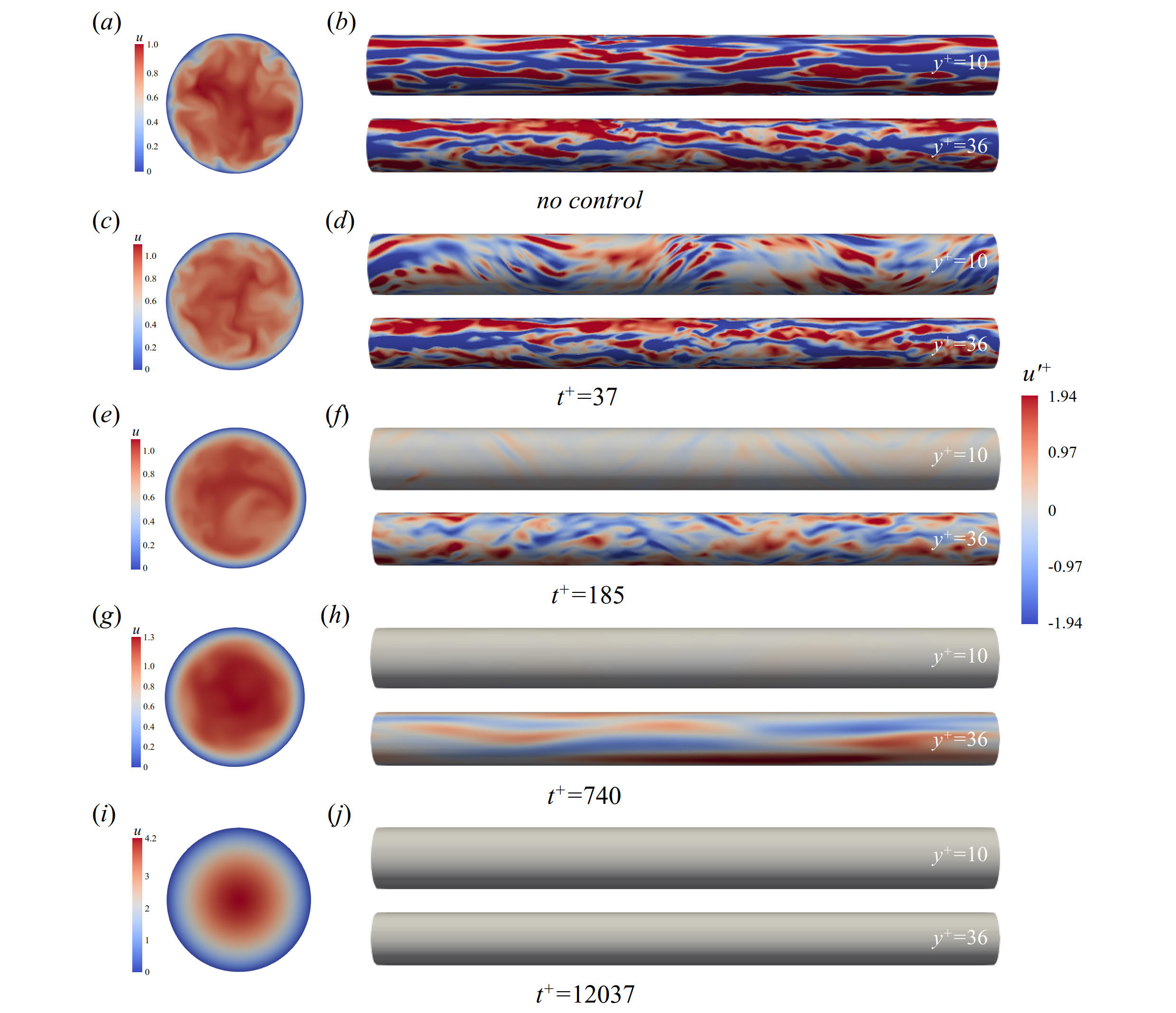}
\caption{\label{fig:instanfield_relminar} Temporal evolution of instantaneous flow fields. Five instants, labelled in the figure, are chosen for presentation. Control is imposed at $t^+$=0.  ($\emph{a}$,$\emph{c}$,$\emph{e}$,$\emph{g}$,$\emph{i}$) $u$ contour of the same cross sectional slice. ($\emph{b}$,$\emph{d}$,$\emph{f}$,$\emph{h}$,$\emph{j}$) contour of streamwise velocity fluctuation $u'$ at two cylindrical surfaces, i.e. $y^+$=10 and 36. The contours on the right column shares the same color legend.}
\end{figure}

Figure \ref{fig:instanfield_relminar} presents the instantaneous flow fields at five instants during the relaminarization. On the left column, the instantaneous $u$ contour of the same cross-section slice are shown,  while on the right column, instantaneous streamwise velocity fluctuations at two cylindrical surfaces, i.e. $y^+$=10 and 36, are presented. These two wall-normal locations are chosen  deliberately based on the largest thickness of SSL. The smaller one is located within the SSL and the larger one is outside the SSL. 

Without control, several tentacle-like high-momentum flow structures, which extend from the core of pipe towards the wall, together with the near-wall low-momentum fluids jigsawed along the azimuthal direction can be clearly observed (figure \ref{fig:instanfield_relminar}(\emph{a})). When control is imposed, a time interval of $t^+$=37 is large enough to allow the near-wall streaks to response to the wall rotation. At this moment, streaks within the SSL incline sinuously while the outer streaks remain nearly straight (figure \ref{fig:instanfield_relminar}(\emph{d})). No remarkable change can be observed from the cross-sectional contour (figure \ref{fig:instanfield_relminar}(\emph{c})). At $t^+$=185, an annular layer can be clearly observed between the core turbulence and the wall (figure \ref{fig:instanfield_relminar}(\emph{e})). Within the layer, the velocity fluctuations are indistinguishable (figure \ref{fig:instanfield_relminar}(\emph{f})), suggesting that the SSL seems to smear out the turbulent motions and produces a local laminar annular region. Meanwhile, the annular SSL blocks the protrusion of high-momentum fluids from the core towards the wall and encloses the core turbulence. Also, the outer turbulence intensity is attenuated (figure \ref{fig:instanfield_relminar}(\emph{f})). With time elapsing, the turbulence in the core region gradually decays  (figure \ref{fig:instanfield_relminar}(\emph{g,h})) and finally the flow reaches a laminar state (figure \ref{fig:instanfield_relminar}(\emph{i,j})). Note that the formation of laminar SSL occurs in a short time scale while the relaxation of core turbulence requires a tremendous amount of time. This is in consonance with figure \ref{fig:validation}(\emph{b}), in which the total TKE falls exponentially at the beginning and then declines very slowly.  

The above qualitatively described process can be quantified by examining the time evolution of the wall-normal profile of TKE, as shown in figure \ref{fig:TKE_grad}(\emph{a}). Indeed, the TKE in the near-wall region declines sharply at the initial stage (0$\textless$$t^+$$\textless$278), followed by the decay of core turbulence since its energy mainly originates from the diffusion process of turbulent activities in buffer layer. At $t^+$=278, the TKE drops to nearly zero in the region of 0$\textless$$y^+$$\textless$20 and peaks at $y^+$$\approx$40. This period corresponds to the formation of laminar SSL. After that, the TKE profile lingers for a while ($t^+$=278-556), and during this time interval, the TKE in the central region is further attenuated. After this short lingering time, the TKE keeps declining, with the peak position moving towards the central region.   

Obviously, two evolution stages can be roughly identified. The first is the formation of laminar SSL and the second is the decay of outer turbulence after the complete formation of laminar SSL. For the first stage, the mechanism that eliminating turbulence within the SSL can be attributed to the enhancement of turbulent dissipation ($\epsilon_k$). This mechanism has been explained by \cite{ricco2012changes}, in which they concluded that the inclination of near-wall streaks enhances the turbulent enstrophy, whose volume-integrated value equals to the total turbulent viscous dissipation. This is valid for present study and confirmed in figure \ref{fig:TKE_pd}(\emph{a}). As can be seen, after the control is imposed, the turbulent dissipation is enhanced at around $y^+$=10 initially, where the near-wall streaks populates. After that, it declines rapidly. Note that, during the first stage, the dissipation rate and production rate ($P_k$) of TKE both declines, but the ratio of production to dissipation quickly decreases below one across the whole radius (see figure \ref{fig:ratio_pre}(\emph{a})), suggesting the continuous decline of TKE. 

The end of first stage is marked by the complete formation of laminar SSL. However, the formation of laminar SSL is accompanied by the decline of velocity gradient at the wall (see figure \ref{fig:TKE_grad}(\emph{b})), which implies that larger part of near-wall regions are affected as the thickness of SSL increases accordingly as noted before. Due to the CPG constraint, the fluids in the outer region accelerates, resulting in a substantial increase of velocity gradient in the buffer layer, as shown in figure \ref{fig:TKE_grad}(\emph{b}). Since $\widetilde{w}$ is much small in the buffer layer, the dominating term that is contributing to the TKE production is $\widetilde{u'v'} \partial \widetilde{u}/ \partial r$. However, the increase of $\partial \widetilde{u}/ \partial r$ could not lead to the increase of TKE production. Instead, it declines continuously (see figure \ref{fig:TKE_pd}(\emph{b})), which means that the shear stress $\widetilde{u'v'}$ declines more rapidly. Nevertheless, according to figure \ref{fig:ratio_pre}(\emph{a}), the ratio of production to dissipation keeps increasing and exceeds one in the region of 30$\textless$$y^+$$\textless$70 during the time interval of $t^+$=278-741 and the peak location of the ratio matches well with that of TKE. This explains why the TKE lingers in this time interval.

It is known that the Reynolds shear stress $\widetilde{u'v'}$ is associated with the near-wall ejection and sweep events. Under the wall-normal gradient of mean velocity, fluids is lifted away from the wall (lift-up effect), forming the low-speed region, and in the meantime, high-momentum fluids in the outer layer sweep downwards (splatting) due to the mass conservation. The rapidly-declined shear stress implies the sharply weakened wall-normal lift-up and splatting events, in which the latter could be quantitatively assessed by examining the pressure-strain term in $\widetilde{v'v'}$ transport equation. As shown in figure \ref{fig:ratio_pre}(\emph{b}), the splatting events weaken rapidly at the boarder of SSL, despite the considerable increase of velocity gradient in that region. 

So far, we have presented the transient dynamics during the relaminarization process in detail. The formation of laminar SSL is the first stage. After that, the continuous attenuation of turbulence activities near the boarder of SSL, which is termed as the second stage, leads to the final relaminarization. However, the mechanism behind the second stage is still unclear. Since the velocity gradient increases significantly, the intensity of turbulent motions is supposed to be strengthened rather than attenuated. To answer this question, we made a comparison between channel and pipe under the same control parameter of case 2 at the same Reynolds number, and the results will be presented in the next section.  

\subsection{Comparison with channel flow}\label{comparison}

\begin{figure}
\centering
\includegraphics[scale=0.65]{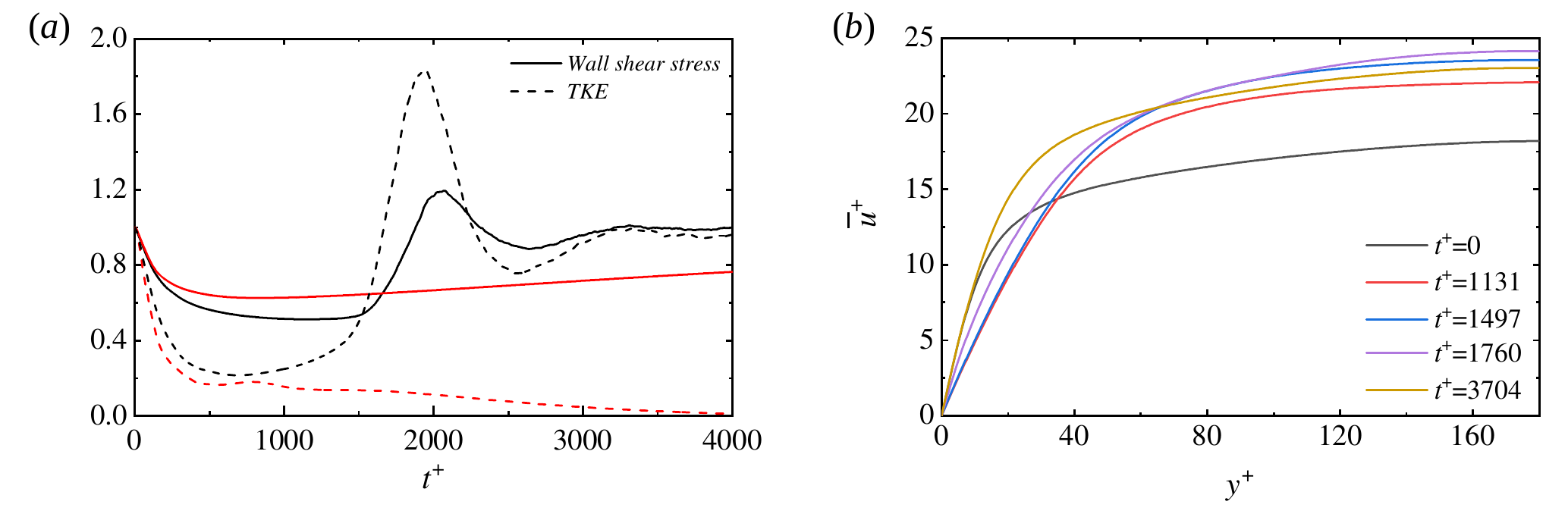}
\caption{\label{fig:channelva_meanvel} (\emph{a}) Temporal evolution of integrated wall shear stress (solid lines) and total TKE (dashed lines). Black lines: channel; Red lines: pipe. All quantities are normalized by the average values of the uncontrolled case. (\emph{b}) Time evolution of streamwise mean velocity profile $\overline{u}$($\emph{y}^+$) in channel.}
\end{figure}

We performed a DNS of turbulent channel flow at Reynolds number of $Re_\tau$=180 with the control parameter same as case 2, i.e. ($\lambda^+$,$A^+$)=(1695,12). The choice of this control parameter is rooted in the fact that we can compare our simulation results with previous studies (\citet{quadrio2011laminar},\citet{viotti2009streamwise}) to ensure the reliability of our results. For the computational details and comparison with previous studies, the readers can refer to Appendix \ref{appA}. Figure \ref{fig:channelva_meanvel}(\emph{a}) compares the temporal evolution of total TKE as well as the integrated streamwise wall shear stress between channel and pipe. In channel, as expected, the flow does not relaminarize in the end. Both the TKE and wall shear stress exhibit the similar trend, that is, they decline initially and reaches its minimum, after which they bounce back to a level that overshoot the averaged value for unactuated state and then drop to its new statistically-steady value. The initial decline for both TKE and wall shear stress are consistent with \citet{ricco2012changes}, in which the channel flow under the temporal wall oscillation control is investigated in the physical constraint of CPG. Nevertheless, the subsequent overshoot is not reported in \citet{ricco2012changes}. This may characterises the difference between the temporal and spatial wall oscillation, which is, however, beyond the scope of this paper. Also, the mean velocity profile (figure \ref{fig:channelva_meanvel}(\emph{b})) exhibits the same trend as pipe flow, that is, in the vicinity of the wall, it decreases initially and then increases to the final value. The implication behind is that the velocity gradient continuously increases in the buffer region. Here, we focus on the transient behaviours before the new fully-developed turbulence state is reached, in an effort to figure out why the flow does not relaminarize.  

\begin{figure}
\centering
\includegraphics[scale=0.65]{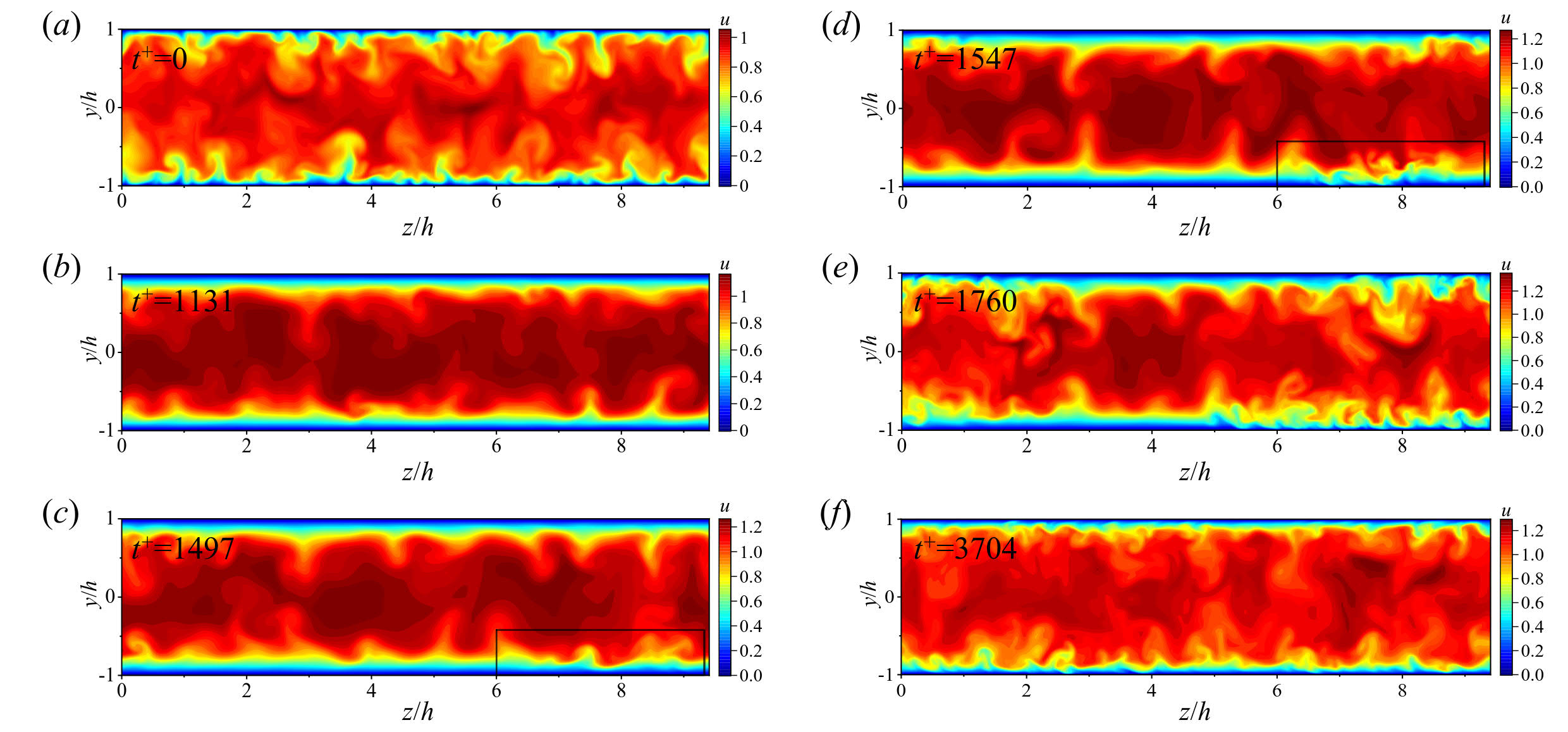}
\caption{\label{fig:slice_channel} Time evolution of instantaneous streamwise velocity contour at a single cross section for channel flow. Control is imposed at $t^+$=0. The enlarged view of black box areas in (\emph{c})(\emph{d}) are shown in figure \ref{fig:enlarge}. }
\end{figure}

\begin{figure}
\centering
\includegraphics[scale=0.65]{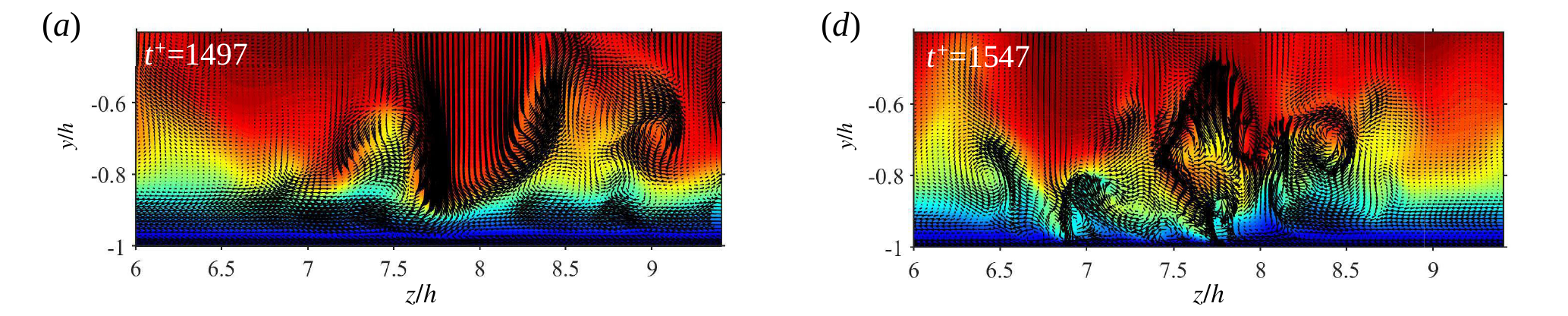}
\caption{\label{fig:enlarge} Enlarged view of black box areas in figure \ref{fig:slice_channel}(\emph{c})(\emph{d}). The black arrows denote the direction of in-plane velocity. The length of arrows represents the magnitude of local velocity.}
\end{figure}

Similar to the pipe flow, firstly, we examine the temporal evolution of flow field at a single cross-sectional slice in channel, shown in figure \ref{fig:slice_channel}. Six snapshots are chosen for examination, with the corresponding time labelled in figures. Moreover, in order to provide more detailed information during the transient period, the enlarged view of black box in figure \ref{fig:slice_channel} is presented in figure \ref{fig:enlarge}, superimposed with the vectors that denote the direction of in-plane local velocity. During the TKE-declined stage, the evolution of flow field in channel is analogous with that in pipe, which is characterized by the formation of laminar SSL and the attenuation of outer turbulence activities. However, as time elapsing, the turbulence outside the SSL would not completely decay. Instead, the downward motion of high-momentum fluids starts to penetrate the laminar SSL (see figure \ref{fig:slice_channel}(\emph{c}) and figure \ref{fig:enlarge}(\emph{a})), which causes the local instability of laminar SSL. As evident in figure \ref{fig:slice_channel}(\emph{d}) and figure \ref{fig:enlarge}(\emph{b}), in the unstable region, the vortices penetrate into the laminar SSL, and the strong transverse convection tilts the ejection and sweep events. Note that the velocity magnitude (length of the vector) in the unstable region is considerably larger than that in surrounding region. Next, the local unstable region contaminates the surrounding laminar SSL (figure \ref{fig:slice_channel}(\emph{e})) and eventually the whole SSL becomes turbulent (figure \ref{fig:slice_channel}(\emph{f})). Hence, the ascent stage of the wall shear stress in figure \ref{fig:channelva_meanvel} can be considered as the contaminating stage. When the instability occurs, the intense generation of vortices increases the impact of fluids against the wall, resulting in the overshoot of integrated wall shear stress. After the transient, the integrated wall shear stress returns to its unactuated value due to the CPG.

\begin{figure}
\centering
\includegraphics[scale=0.65]{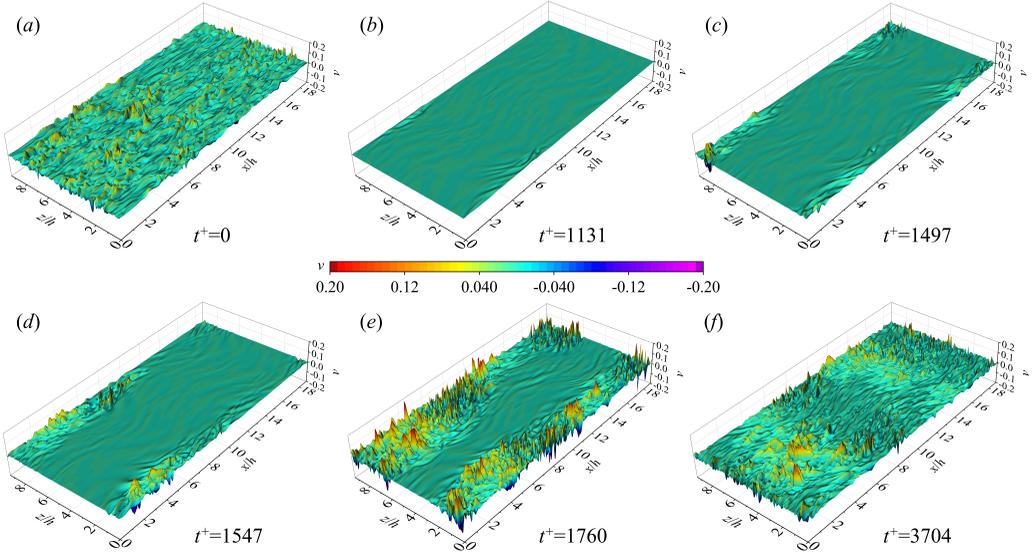}
\caption{\label{fig:xslice} Time evolution of instantaneous three-dimensional $v$ contour at $y^+$=10 for channel.}
\end{figure}

Since the fluids convect downstream, the time evolution of flow field at a fixed cross section could not fully exhibit the evolution of flow field. Hence, we further present the temporal evolution of instantaneous three-dimensional $v$ contour in the $x$-$z$ plane at $y^+$=10, as shown in figure \ref{fig:xslice}. The transition of $v$ contour from a rugged surface, which represents the random upward and downward motion, to smooth flat surface clearly indicates the formation of laminar SSL. After that, local fluctuations occur, contaminating the surrounding laminar region, and in the meantime, it convects downstream and expands. In the local fluctuating region, the peak of $v$ contour is evidently larger and the number of peaks also increases, which corresponds to the transient overshoot of wall shear stress. Eventually, the whole SSL becomes turbulent and the streamwise wavy pattern can be clearly observed.

\begin{figure}
\centering
\includegraphics[scale=0.65]{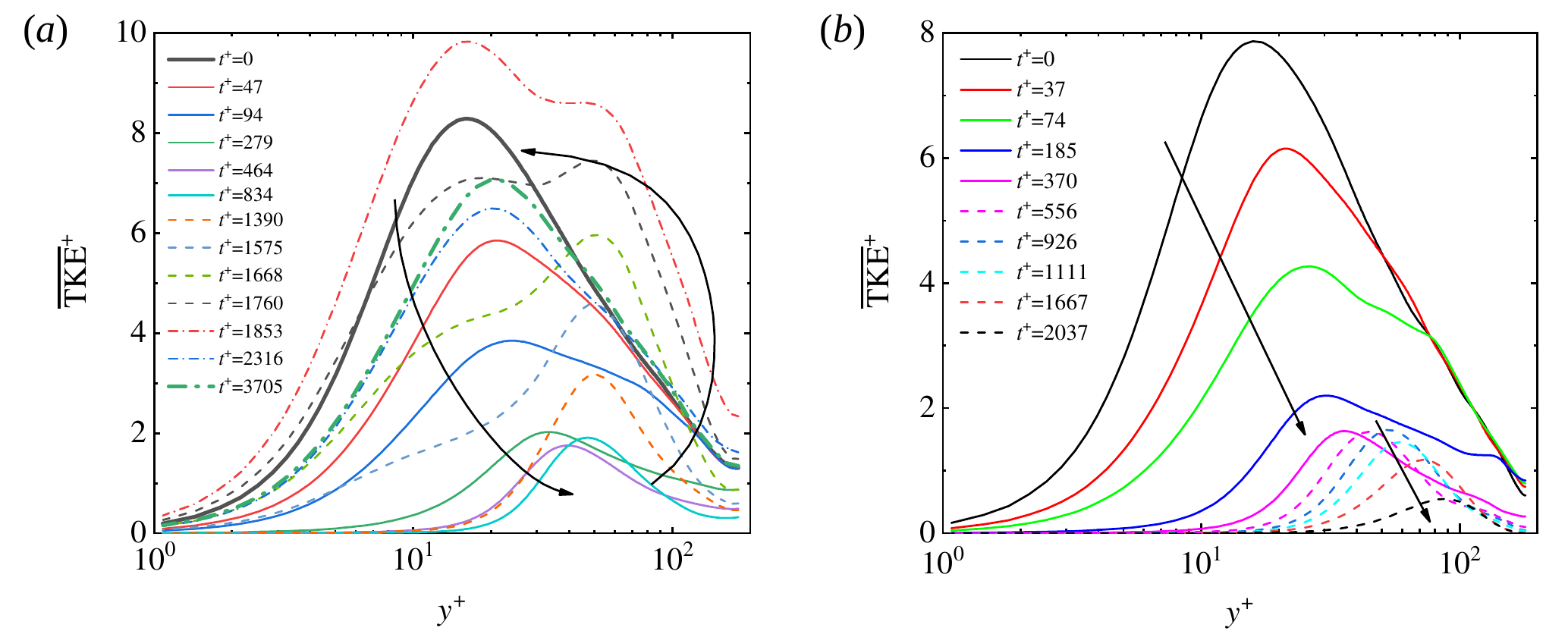}
\caption{\label{fig:channel_pipe_tke} Time evolution of wall-normal profile of normalized TKE. (\emph{a}) channel. (\emph{b}) pipe. The arrows denote the direction of change in time.}
\end{figure}

\begin{figure}
\centering
\includegraphics[scale=0.65]{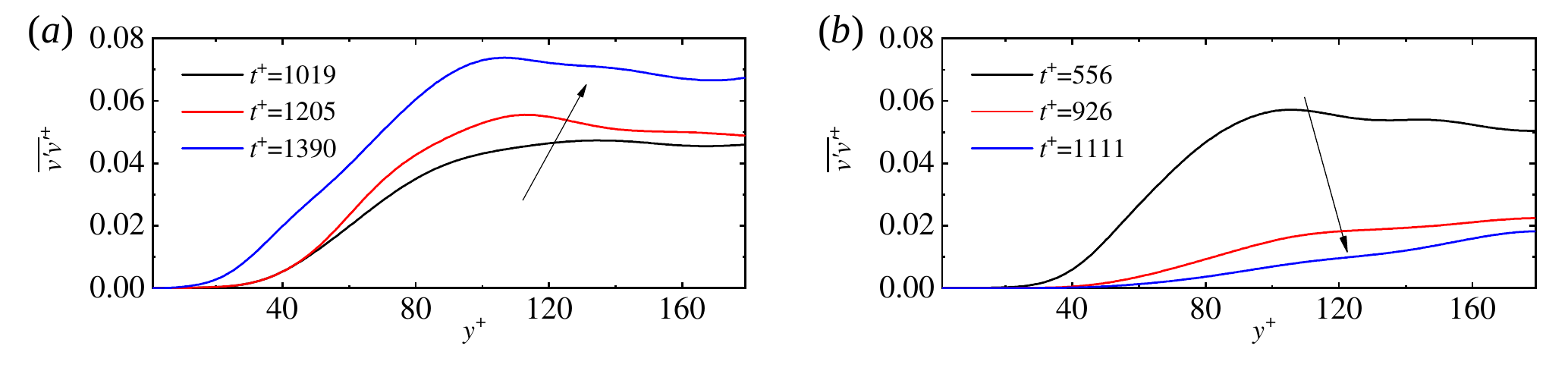}
\caption{\label{fig:channel_pipe_vv} Time evolution of wall-normal profile of non-dimensionalized wall-normal stress $\overline{v'v'}$. (\emph{a}) channel. (\emph{b}) pipe. The arrows denote the direction of change in time. The time interval chosen for channel corresponds to the initial stage of the TKE bounce-back. For pipe, the time interval corresponds to the decay of outer turbulence. }
\end{figure}

Figure \ref{fig:channel_pipe_tke} compares the temporal evolution of wall-normal TKE profiles for channel and pipe flow under the same control parameters, i.e. ($\lambda^+$,$A^+$)=(1695,12). As can be seen, in the TKE-declined stage, the flow evolution for channel is consistent with pipe, which is characterised by the formation of laminar SSL. The crucial difference is that the laminar SSL in channel will eventually be disrupted, while in pipe, it remains stable. This is reflected by the fact that the TKE bounces back in channel and decays completely in pipe. Note that the velocity gradient also keeps increasing in the buffer region in channel, hence we speculate that the increase of velocity gradient enhances the turbulent motion outside the SSL, strengthening the impact of external turbulence on the laminar SSL and finally the laminar SSL loses its stability when the magnitude of such impact is beyond some certain threshold. The scenario can be simply confirmed by examining the temporal evolution of wall-normal stress $\overline{v'v'}$, shown in figure \ref{fig:channel_pipe_vv}, since $\widetilde{v'v'}$ dictates the production of $\widetilde{u'v'}$ and hence governing the production of TKE outside the SSL. As expected, $\overline{v'v'}$ is continuously enhanced after the laminar SSL is completely formed in channel, which is completely opposite to that in pipe where $\widetilde{v'v'}$ decreases rapidly, indicating that the impact of outer turbulent motions on laminar SSL is gradually enhanced. 

By comparing the channel and pipe flow, we know that under the circumstance of continuously increasing velocity gradient, the impact of outer turbulent motions on laminar SSL in channel is continuously enhanced. Whereas in pipe, it is continuously attenuated. We believe that finding the origin of such difference is crucial to understand the physical mechanism that leads to the relaminarization in pipe flow, which will be presented in the next section.

\subsection{Relaminarization mechanism}\label{mechanism}

It is known, from the previous section, that the relaminarization in pipe is closely related to the stability of annular SSL, hence we proceed to find the mechanism behind in terms of energy flux that pertains to circumferential mean flow in pipe. As shown in equation (\ref{equ:cir_mean_flow}), the energy input eventually flows into two places, one is the mean dissipation, which converts the energy into heat, and the other is energy exchange with TKE. In $\overline{w'w'}$ transport equation, the leftmost two terms appear as the production, which means Reynolds stress $\overline{w'w'}$ drains energy from circumferential mean flow. Term $\wp$ is of great interest, it appears both in the transport equation of $\overline{w'w'}$ and $\overline{v'v'}$, which reads:

$\overline{\frac{1}{2}v'v'}$ budget:
\begin{equation}\label{vv}
\frac{{\partial \overline {\frac{1}{2}v'v'} }}{{\partial t}} = {D_v} + {V_v} + {\varepsilon _v} + 2\underbrace{\frac{{\overline {\widetilde {v'w'}\widetilde w} }}{r}}_\textsc{$\wp$}.
\end{equation}

$\overline{\frac{1}{2}w'w'}$ budget:
\begin{equation}\label{ww}
\frac{{\partial \overline {\frac{1}{2}w'w'} }}{{\partial t}} = {D_w} + {V_w} + {\varepsilon _w} + {P_w}-\underbrace{\frac{{\overline {\widetilde {v'w'}\widetilde w} }}{r}}_\textsc{$\wp$},
\end{equation}
where the production rate $P_w$ is:
\begin{equation}
{P_w} = -\overline {\widetilde {v'w'}\frac{{\partial \widetilde w}}{{\partial r}}}  - \overline {\widetilde {u'w'}\frac{{\partial \widetilde w}}{{\partial x}}}.
\end{equation}
Equation (\ref{equ:cir_mean_flow}) can be reorganized as follow (time average is excluded):
\begin{equation}\label{cir_mean_2}
\frac{{\partial \overline {\frac{1}{2}\widetilde w\widetilde w} }}{{\partial t}} = {E_w} + {\phi _w} - {P_w} - \underbrace{\frac{{\overline {\widetilde {v'w'}\widetilde w} }}{r}}_\textsc{$\wp$}.
\end{equation}
In the above equations, $D$ denotes the diffusion term, $V$ represents the velocity-pressure gradient term, $\epsilon$ is the turbulent dissipation rate, $E$ is associated with the energy input and $\phi$ corresponds to the mean dissipation rate. For a complete expression of the above equations, the readers can refer to the Appendix \ref{appB}.  

\begin{figure}
\centering
\includegraphics[scale=0.65]{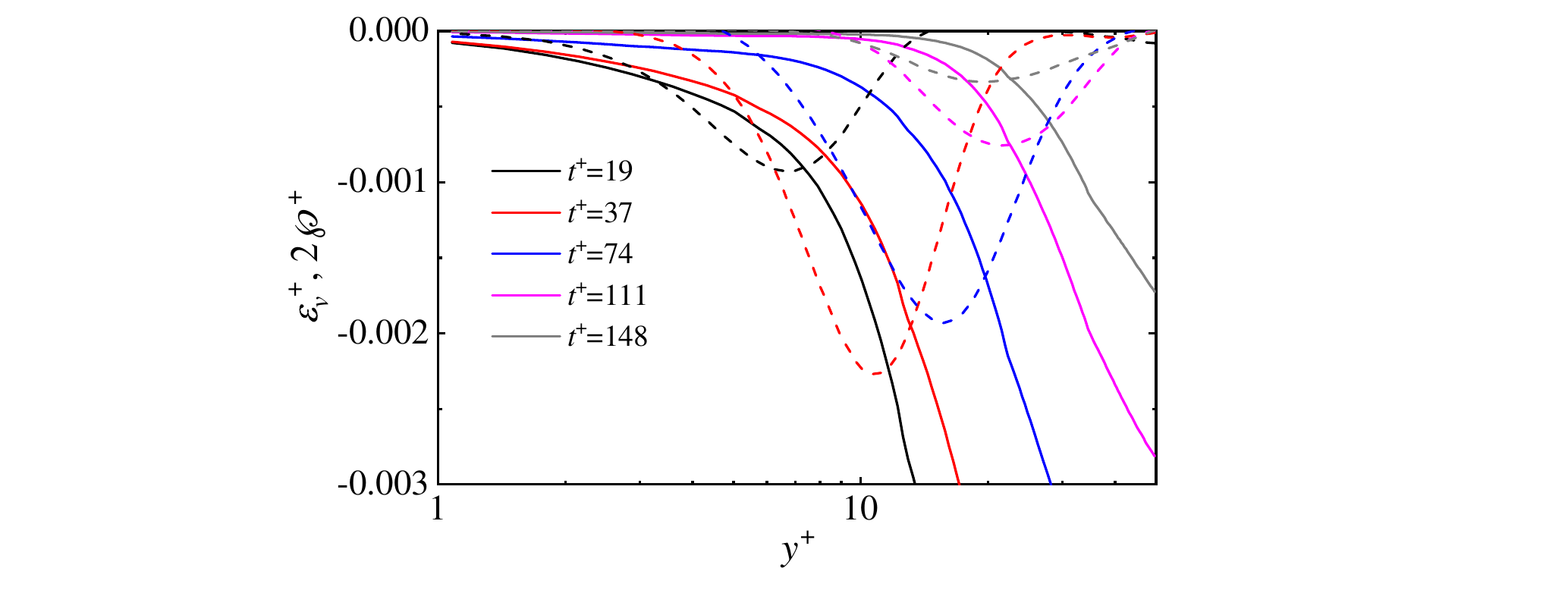}
\caption{\label{fig:dissi_extra} Time evolution of wall-normal profile of $\epsilon_v^+$ (solid lines) and term 2$\wp^+$ (dashed lines) for case 5 at initial stage.}
\end{figure}

\begin{figure}
\centering
\includegraphics[scale=0.65]{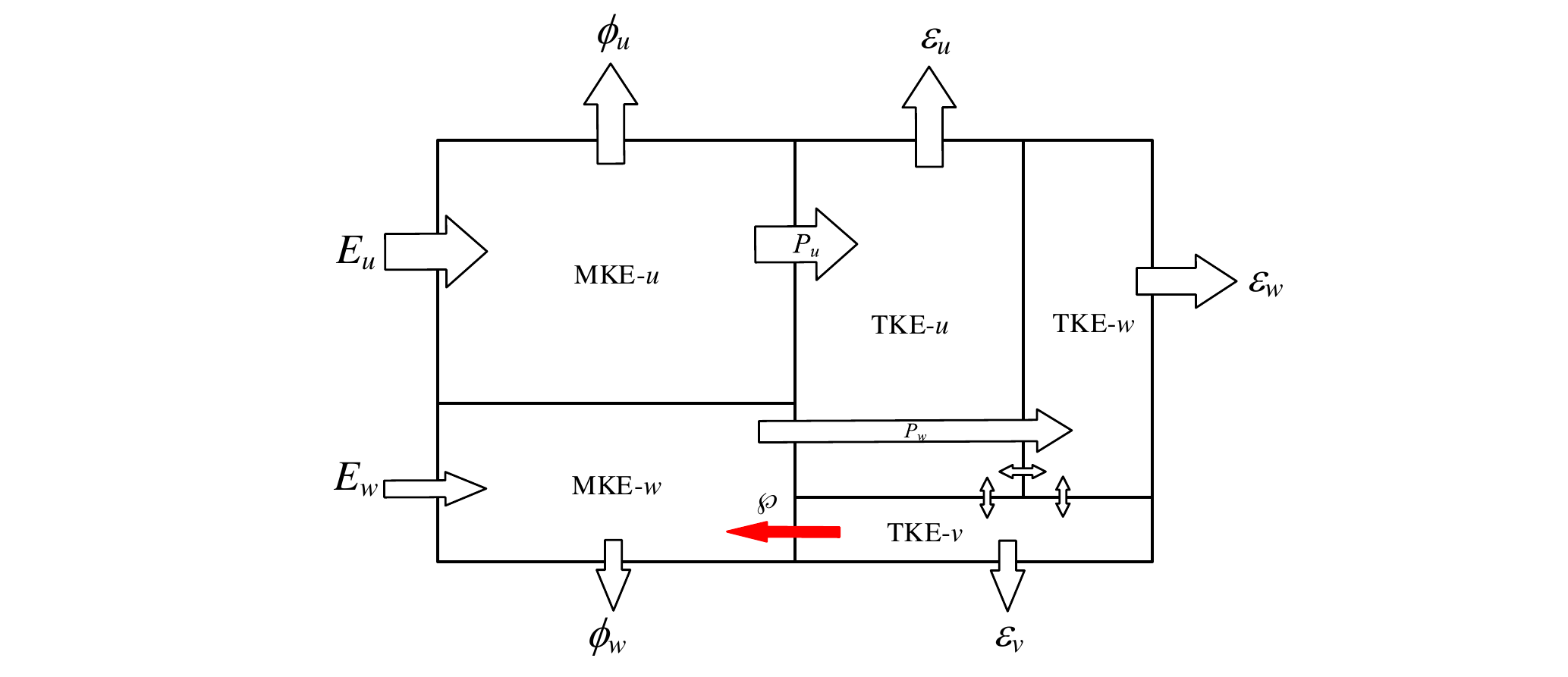}
\caption{\label{fig:energy box} Energy box for the controlled pipe flow. $E$, $\phi$, $P$ and $\epsilon$ denotes the energy input, mean dissipation, turbulent production and turbulent dissipation respectively. The subscript denotes the velocity components that energy pertains to. The red arrow highlights the energy transfer from TKE-$v$ to MKE-$w$, namely term $\wp$, which is absent in channel.}
\end{figure}

We first examine the role of term $\wp$ in $\overline{v'v'}$ budget (\ref{vv}). Obviously, only when the SSL is turbulent can the effect of term $\wp$ appears. Hence, figure \ref{fig:dissi_extra} presents the time evolution of wall-normal profiles of term 2$\wp$ and $\epsilon_v$ for case 5 at the initial stage after the control is imposed. It can be observed that term $\wp$ acts as a sink term and is comparable to $\epsilon_v$ in the region of $y^+$=10-30. After the control is imposed, $\epsilon_v$ declines monotonically while term $\wp$ increases initially and reaches its maximum at around $t^+$=37, after which it vanishes gradually, accompanied with the peak moving away from the wall. Note that the sign of term $\wp$ in equation (\ref{ww}) and (\ref{cir_mean_2}) is opposite to that in equation (\ref{vv}),
which means that the role of term $\wp$ is draining energy from $\overline{v'v'}$ and redistribute equally to the circumferential mean flow and $\overline{w'w'}$. The latter part of energy pertains to the inter-component exchange of TKE. The portion of energy flowing into the circumferential mean flow is of great interest as it indicates a new path of energy exchange between the mean flow and turbulence. Thus, following the procedure introduced in \citet{ricco2012changes}, the global energy balance in pipe flow can be summarised in figure \ref{fig:energy box}. Although the streamwise variation of wall velocity implies that $\widetilde{v}$ is not zero, but in fact, $\widetilde{v}$ is examined to be much smaller compared with $\widetilde{u}$ and $\widetilde{w}$ . Hence, the MKE (mean kinetic energy) that pertains to radial direction is omitted here. The red arrow highlights the energy transfer from TKE-$v$ to MKE-$w$, namely the term $\wp$, that is \emph{absent} in channel flow. Clearly, different from the scenario in channel where the spanwise mean flow only feed energy into TKE and the balance of TKE is attained only by production and dissipation rate (\citet{ricco2012changes}), in pipe flow, there also exists the energy flowing from TKE, specifically the TKE in radial direction, into the circumferential mean flow, highlighting the difference between channel and pipe. It is worth mentioning that the global energy flux associated with term $\wp$ might be negligible when compared with global TKE production or dissipation since term $\wp$ is only confined in the very near-wall region. But as shown in figure \ref{fig:dissi_extra}, term $\wp$ dominates the $\overline{v'v'}$ budget in the very near-wall region (around $y^+$=10), implying that term $\wp$ plays a crucial role in the attenuation of $\overline{v'v'}$. Note that $\overline{v'v'}$ corresponds to the near-wall wall-normal turbulent motions (anti-splatting or splatting) and is crucial to the turbulent self-sustaining mechanisms.

Now we proceed to discuss the physical meaning of term $\wp$. In fact, $\widetilde{w}$/$r$ is the local angular velocity, so the role of term $\wp$ can be interpreted as follow: inside the SSL where the circumferential velocity is large enough, the interaction between strong rotation and stress $\widetilde{v'w'}$ continuously drains energy from wall-normal stress $\overline{v'v'}$ into the circumferential mean flow, leading to the continuously decline of $\overline{v'v'}$. The physical implication is that due to the term $\wp$, any wall-normal turbulent motions, specifically the wall-normal splatting and lift-up events, will be absorbed by the annular SSL in pipe, which results in the disruption of turbulence self-sustaining mechanisms. In other words, the annular laminar SSL is stable because it can continuously absorb energy from outer turbulence, leading to the decay of outer turbulence and the final relaminarization.  Whereas in channel, the turbulence self-sustaining mechanism could be maintained since the spanwise mean flow would not absorb energy from wall-normal stress. As the flow rate increases, the impacts of outer turbulent motions on laminar SSL are enhanced due to the continuously increasing velocity gradient. When the strength of such impact increases beyond some certain threshold, the transition of flat SSL from laminar state to turbulent state will be triggered and the whole flow system becomes turbulent.  

\begin{figure}
\centering
\includegraphics[scale=0.65]{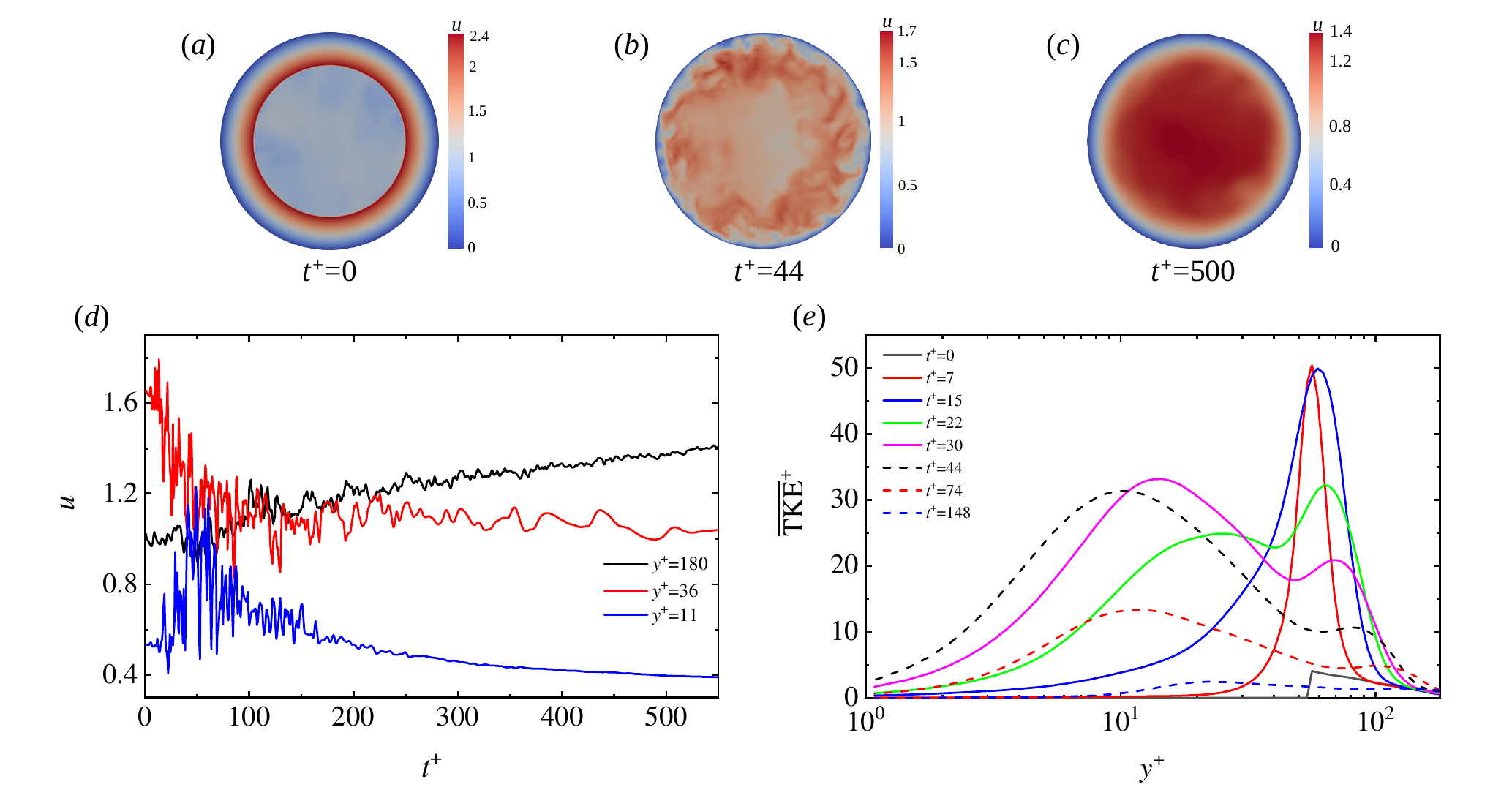}
\caption{\label{fig:per169530} DNS of case 5 with intense perturbation imposed at initial fully-developed laminar field. (\emph{a}) The initial composite flow field. (\emph{b}) Instantaneous flow field at $t^+$=44 where the laminar SSL is turbulent. (\emph{c}) Instantaneous flow field at $t^+$=500 where the turbulence is decaying. (\emph{d}) Temporal evolution of streamwise velocity at $y^+$=11,36,180. (\emph{e}) Temporal evolution of wall-normal profile of TKE. }
\end{figure}

Next we further demonstrate that the laminar SSL in pipe is stable at $Re_\tau$=180 if the control wavelength and velocity amplitude are both large enough. Case 5 is selected as the object. In the fully-developed laminar state, we chose a single flow field and replace the data in the region of 0$\textless$$r$$\textless$0.35 by the instantaneous velocity and pressure fields extracted from a single fully-developed turbulent flow field. The $u$ contour of cross-sectional slice of the composite flow field is shown in figure \ref{fig:per169530}(\emph{a}). Note that $r$=0.35 ($y^+$=54) is well beyond the SSL and we take this composite flow field as the starting point for the subsequent simulation. We note that this approach has been employed in \citet{wu2015osborne} to study the gradual transition from laminar to fully-developed turbulence in pipe flow. The purpose of this simulation is to verify whether the laminar SSL is stable to such intense perturbations.  

\begin{figure}
\centering
\includegraphics[scale=0.65]{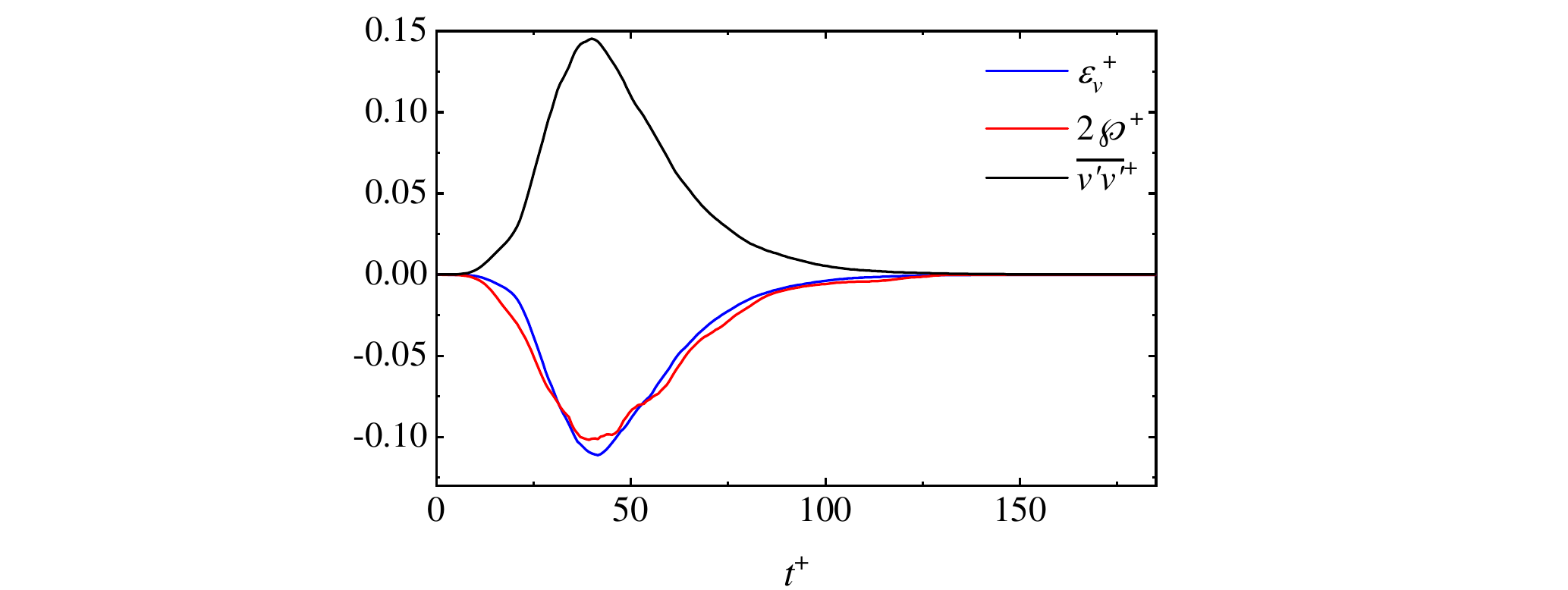}
\caption{\label{fig:per169530y7} Time evolution of $\epsilon_v^+$, 2$\wp^+$ and $\overline{v'v'}$ at $y^+$=7 during the transient in the DNS of case 5 where the intial field is imposed with intense perturbation. Note that $\overline{v'v'}$ is divided by a factor of 10.}
\end{figure}

The time evolution of streamwise velocity at different wall-normal positions and of wall-normal profiles of TKE are shown in figure \ref{fig:per169530}(\emph{d}) and figure \ref{fig:per169530}(\emph{e}) respectively, together with the instantaneous $u$ contour of cross-sectional slice at three instants shown in figure \ref{fig:per169530}(\emph{a}-\emph{c}). At the beginning, the sudden variation of streamwise velocity at $r$=0.35 causes the intense transient turbulence burst. The TKE increases dramatically in this region, reaching the maximum at $t^+$=7-15. After that, the laminar SSL is disturbed, which is reflected by the rapid growth of TKE below $y^+$=40 and the strong fluctuation of streamwise velocity at $y^+$=11 (figure \ref{fig:per169530}(\emph{a})). This is also visually confirmed by the snapshot at $t^+$=44 in figure \ref{fig:per169530}(\emph{b}). Besides, the disruption of laminar SSL is accompanied by the relaxation of outer turbulence after the transient burst. At around $t^+$=30, the TKE inside the SSL reaches the maximum, suggesting the complete disruption of laminar state. As time elapsing, surprisingly however, the strong velocity fluctuations gradually decay and the velocity in the central region rises smoothly. The wall-normal profile of TKE at $t^+$=148 clearly indicate that the flow returns back to laminar state. This results is interesting as it demonstrates the stability of laminar SSL in pipe at this Reynolds number. Even such high magnitude of perturbations can only temporarily disturb the laminar SSL, and eventually the system will return back to laminar state. The implication is that, from the perspective of system dynamics, under the control parameter of case 5, the laminar state is probably the only fixed point for pipe flow at this Reynolds number. We also conduct the same simulation under the control parameter of case 2. The results show that the flow also experiences a transient chaos and returns back to laminar state, which are hence not shown here. Again, we examine the time evolution of 2$\wp$, $\epsilon_v$ in $\overline{v'v'}$ budget and $\overline{v'v'}$ itself at $y^+$=7 during the transient burst for case 5. As shown in figure \ref{fig:per169530y7}, term $\wp$ is comparable to $\epsilon_v$, which strongly suggests that term $\wp$ plays an important role in the attenuation of wall-normal stress. Also, the increase of term $\wp$ is accompanied by the increasing of turbulence intensity, indicating that the effect of term $\wp$ is self-adapting, that is, the larger the turbulence intensity, the more energy absorbed by circumferential mean flow, which in turn resulting in the decline of turbulence intensity.    

Tough we claim that the annular SSL can absorb energy from wall-normal stress, it is hard to find distinguishable change if we examine the time evolution of global energy that pertains to circumferential mean flow since the energy of wall-normal stress is too small compared with circumferential mean flow. So we reiterate here that term $\wp$ contributes dominantly to the attenuation of wall-normal stress within the SSL, which causes the subsequent destablization of turbulence. After highlighting the importance of term $\wp$, it is easy to understand why it requires simultaneously large control wavelength and large velocity amplitude to relaminarize the turbulence in pipe flow. Term $\wp$ is the product of local Reynolds stress $\widetilde{v'w'}$ and local angular velocity. As already shown in figure \ref{fig:spatial stokes layer2}, large control wavelength corresponds to large thickness of SSL, hence producing large value of Reynolds stress $\widetilde{v'w'}$. On the other hand, large velocity amplitude generates large angular velocity. If either condition is not satisfied, the role of term $\wp$ would be negligible and the flow could not be relaminarized. Another issue is about the effect of Reynolds number. The above conclusions are drawn in the context of low Reynolds number of $Re_\tau$=180, whether it still holds at high Reynolds number remains unclear. At high Reynolds number, the low thickness of boundary layer and the modulation of near-wall turbulence by outer large-scale motions would undoubtedly influence the control results, which is our future research. 

\section{Summary}\label{summary}

In this study, we have examined the effects of streamwise-varying wall rotation on turbulent pipe flow by DNS at Reynolds number $Re_\tau$=180. Two control parameters, which are velocity amplitude and wavelength, are considered. When one of the control parameter is small enough, the flow remains turbulent and drag reduction is achieved, which manifests itself as the increase of mass flow rate due to the physical constraint of CPG employed in present study. When the two control parameters are both large enough, the flow relaminarizes. Whereas in turbulent channel flow at similar Reynolds number, no relaminarization is reported. Such significant difference suggests the important role of geometry when employing this wall-motion-based control strategy. In terms of energetic performance, high amplitude will significantly increase the power input and hence reduce the net energy saving rate. Positive net energy saving is achievable when the control wavelength is large and the amplitude is small. An examination of effectiveness reveals that increasing the wavelength is probably more effective than increasing the amplitude in achieving the drag reduction.

The transverse boundary layer, which is called spatial Stokes layer (SSL), significantly affects the control results. For non-relaminarization cases, two drag reduction scenarios can be identified based on the thickness of SSL. In the case of low thickness, only a small fraction of near-wall streaks are affected and hence the SSL acts as spacer layer lying between the wall and near-wall flow structures, preventing the near-wall wall-normal splatting and thereby reduces the shear stress. An inspection of one-dimensional spectra map reveals that the structures outside the SSL are stretched due to the increased mean velocity gradient. Within the SSL, the strong shearing significantly diminishes the turbulence intensity and the dominant scale of motions is closely related to the control wavelength. When the thickness is large enough to cover more proportion of near-wall streaks, the streaks is inclined sinuously in streamwise direction, forming a wavy pattern and followed by a energy transfer from streamwise stress to circumferential stress. However, the dominant scale of wavy streaks is much less than the control wavelength and even much less than that in uncontrolled case. Such shortening effect is believed to disrupt the original downstream development of near-wall flow structures and whereby Reynolds shear stress is reduced. The wavy pattern of streaks lags behind the direction of local mean velocity because it conforms to the direction of local mean shear stress. 

In terms of the SSL velocity profile, the discrepancy between the present DNS results in turbulent regime and the laminar analytical solution is closely related to the control parameters. With the aid of dimensional analysis on the governing equation, it is shown that high amplitude together with low wavelength tends to increases the magnitude of mean convection term, to which the turbulence modulation could be comparatively small, thereby leading to the small discrepancy. On the contrary, large wavelength and low amplitude will produce large discrepancy. 

For the relaminarization case in pipe, the relaminarization process can be divided into two stages. The first is the formation of laminar SSL, which is caused by the enhancement of turbulence dissipation as discussed in \citet{ricco2012changes} and results in the sharp decline of wall shear stress and turbulence intensity. Due to the constant pressure gradient, the wall-normal gradient of streamwise mean velocity ($\partial \overline{u}$/$\partial r$) increases in the buffer region. The second stage is the decay of turbulence outside the laminar SSL. This stage is characterised by the continuously declined production of TKE despite the fact that $\partial \overline{u}$/$\partial r$ continuously increases. However, the ratio of production to dissipation temporarily exceeds one in the buffer region, which leads to the temporary lingering of TKE. For channel flow under the same control, the first stage is the same as pipe flow. Whereas after the complete formation of laminar SSL, the TKE outside the laminar SSL in channel bounces back rather than completely decay. This is the crucial difference between channel and pipe. For the TKE bounce-back stage, the laminar SSL is disrupted locally at the beginning, and the perturbation gradually contaminates the surrounding laminar region. Finally, the whole laminar SSL loses its stability and the whole flow system returns back to turbulence. It is found that due to the increased $\partial \overline{u}$/$\partial r$, the impact of outer turbulence on laminar SSL is gradually enhanced, which causes the local instability of laminar SSL.  

By examining the energy flux that pertains to circumferential mean flow in pipe, it is found that due to the rotation effect, part of the energy that pertains to wall-normal stress is continuously absorbed by circumferential mean flow, which is believed to be the root cause of continuous decay of outer turbulence since wall-normal stress plays an important role in the turbulence self-sustaining mechanisms. Note that this effect is self-adapting, i.e., the larger the turbulence intensity, the more energy absorbed by circumferential mean flow. Whereas in channel, such effect is absent since no rotation effect exists in the geometry of flat plane. Hence we concluded that the laminar SSL is stable at the Reynolds number considered in present study as it can absorb energy from outer turbulence. This is confirmed by the additional simulations where we manually impose disturbances of great strength to the laminar annular SSL. The fact that the laminar SSL undergoes a transient chaos and returns back to laminar state strongly demonstrates our conclusion.        

\appendix
\section{Details for channel flow simulations}\label{appA}

\begin{figure}
\centering
\includegraphics[scale=0.6]{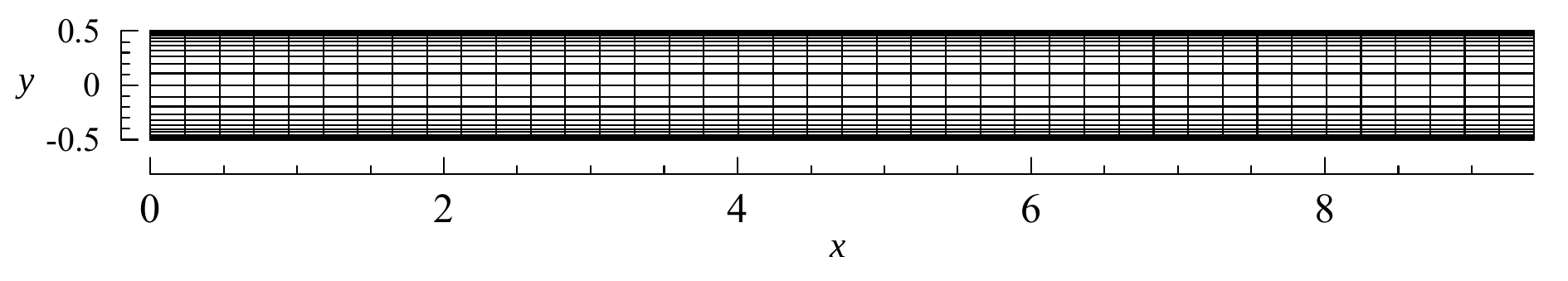}
\caption{\label{fig:channel_mesh} Two-dimensional spectral element meshes for channel flow simulation, with 880 elements in the $x$-$y$ plane.}
\end{figure}

The channel flow DNS at Reynolds number ($Re_\tau$=180) is performed in Cartesian coordinate system using the spectral element-Fourier DNS solver $\emph{semtex}$ (\cite{blackburn2019semtex,blackburn2004formulation}). The simulations is conducted in the physical constraint of CPG and $x$,$y$,$z$ represents the streamwise, wall-normal and spanwise direction respectively. The two-dimensional spectral element mesh, shown in figure \ref{fig:channel_mesh}, is deployed to discretize the $x$-$y$ plane, with 288 ($Nz$) Fourier planes in $z$ direction to represent the three-dimensional computational domain. The element height follows a geometric degression from the centerline to the wall. A 10th-order nodal shape function is employed, which means $P$=11.  The resulting grid resolutions, scaled by the friction velocity, are $\Delta$$x^+$, $\Delta$$y^+_{min/max}$, $\Delta$$z^+$=8.5,0.42-3.9,5.9, covering a channel section of 6$\pi$$h$$\times$3$\pi$$h$$\times$2$h$, where $h$ is the half channel height. Such grid resolutions yield a total computational nodes of approximately 3.1$\times$$10^7$, which is little smaller than that in \citet{quadrio2009streamwise} and \citet{viotti2009streamwise}.  To ensure that the differences between channel and pipe are only caused by geometry, the friction velocity $u_\tau$ and kinematic viscosity $\nu$ are set to be equal to pipe flow. Similarly, a single flow field at fully-developed turbulent state is chosen as the starting point for the subsequent simulation with spatial wall oscillation. 

\begin{figure}
\centering
\includegraphics[scale=0.65]{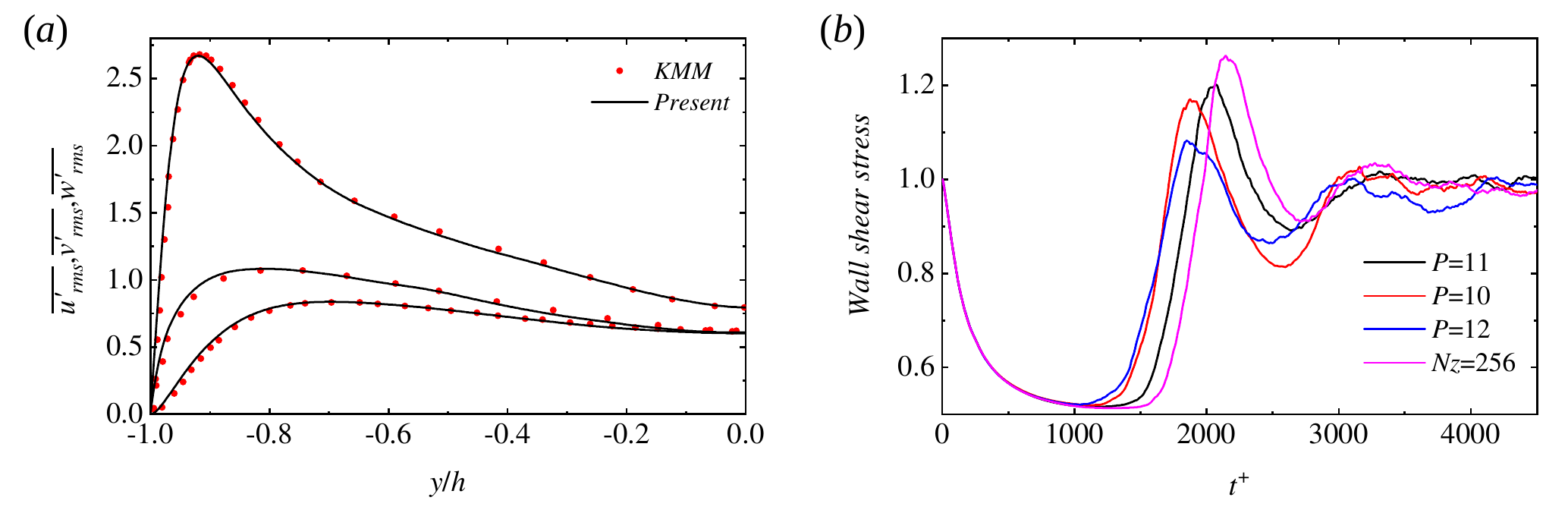}
\caption{\label{fig:channel_validation} Validation of channel flow simulation results. (\emph{a}) Comparison of turbulence intensity in uncontrolled channel between present and \citet{kim1987turbulence} (denoted as KMM). (\emph{b}) Time evolution of total TKE ($\int_0^{3\pi h}$[TKE]d$z$) during the transient in channel for different meshes. The value of total TKE is normalized by the average value of uncontrolled case. The baseline mesh is $P=11$, $N_z$=288. For other three meshes, only one parameter (either $P$ or $N_z$) is varied to with respect to the baseline mesh.}
\end{figure}

The calculated drag reduction for channel flow under the control parameter of ($\lambda^+$,$A^+$)=(1695,12) (same as case 2) is 38.2\%. As shown in figure \ref{fig:drag reduction result}, this result is lower than the drag reduction curve reported in \citet{viotti2009streamwise}, in which the channel flow DNS is conducted in the physical constraint of constant flow rate (CFR). Such difference is reasonable as \citet{quadrio2011laminar} showed that under this control parameter, the drag reduction for CPG is lower than that for CFR. Besides, the present results from uncontrolled channel agrees well with that in \citet{kim1987turbulence} (figure \ref{fig:channel_validation}(\emph{a})), indicating that present mesh is appropriate to simulate the fully-developed turbulence in channel. We also simulate the transient behaviour of controlled channel flow under different mesh resolution (i.e. changing the value of $P$ and $Nz$) to ensure the mesh independence. As shown in figure \ref{fig:channel_validation}(\emph{b}), the initial decline and the subsequent bounce-back and overshoot of wall shear stress occurs in all the simulations with different mesh resolution, indicating that such behaviour is authentic and physically relevant. Moreover, the discrepancy of bounce-back stage for different meshes suggests that the contamination of laminar SSL is sensitive to the grid resolution.    

\section{Reynolds stress transport equation in pipe}\label{appB}

Here we presents the $\frac{1}{2}\widetilde{v'v'}$ and $\frac{1}{2}\widetilde{w'w'}$ transport equation in full details. The mean convection term ($M$) disappears after the streamwise average due to the streamwise periodicity of the control law (\ref{equ:bc}). For the notation $\sim$, time average is excluded.

$\frac{1}{2}\widetilde{v'v'}$ budget:
\begin{equation}\label{equ:vbudget}
\begin{split}
\frac{1}{2} \frac{\partial \widetilde{v'v'}}{\partial t}=\underbrace{-\frac{{\partial r\widetilde {v^{'}v^{'}v^{'}} }}{{2r\partial r}}-\frac{{\partial \widetilde {v^{'}v^{'}u^{'}} }}{2{\partial x}}+\frac{{\widetilde {v^{'}w^{'}w^{'}} }}{r}+ \frac{1}{2} \nu \left( {{\nabla ^2}\widetilde {v^{'}v^{'}}} + \frac{{2\left( {\widetilde {w^{'}w^{'}}  - \widetilde {v^{'}v^{'}} } \right)}}{{{r^2}}} \right)}_\textsc{$D_v$} \\
 \underbrace{-\frac{1}{2}\widetilde {{u}} \frac{{\partial \widetilde {v'v'} }}{{\partial x}}}_{\textsc{$M_v$}} \underbrace{-\widetilde {{v^{'}}\frac{{\partial p^{'}}}{{\partial r}}}}_{\textsc{$V_v$}} \underbrace{-\nu \left[ {\widetilde {{{\left( {\frac{{\partial v^{'}}}{{\partial z}}} \right)}^2}}  + \widetilde {{{\left( {\frac{{\partial v^{'}}}{{\partial r}}} \right)}^2}}  +  \widetilde{\frac{1}{{{r^2}}}{{\left( {\frac{{\partial v^{'}}}{{\partial \theta }} - w^{'}} \right)}^2}} } \right]}_{\textsc{$\epsilon_v$}} + 2\underbrace{\frac{{\widetilde {v^{'}w^{'}} \widetilde {{w}} }}{r}}_{\textsc{$\wp$}}
\end{split}
\end{equation}

$\frac{1}{2}\widetilde{w'w'}$ budget:
\begin{equation}\label{equ:transport3}
\begin{split}
\frac{1}{2} \frac{\partial \widetilde{w'w'}}{\partial t}=\underbrace{-\frac{{\partial r\widetilde {w'w'v'} }}{{2r\partial r}}-\frac{{\widetilde {w'w'v'} }}{r}-\frac{{\partial \widetilde {w'w'u'} }}{{\partial x}}+\frac{1}{2} \nu \left( {{\nabla ^2}\widetilde {w'w'}- \frac{{2\left( {\widetilde {w'w'} - \widetilde {v'v'} } \right)}}{{{r^2}}}} \right)}_{\text{$D_w$}} \\
\underbrace{-\frac{1}{2}\widetilde {{u}} \frac{{\partial \widetilde {w'w'} }}{{\partial x}}}_{\text{$M_w$}} \underbrace{-\widetilde {v'w'} \frac{{\partial \widetilde {{w }} }}{{\partial r}} - \widetilde {u'w'} \frac{{\partial \widetilde {{w}} }}{{\partial x}}
-\frac{{\widetilde {v'w'} \widetilde {{w}} }}{r}}_{\text{$P_w$}}\\
\underbrace{-\frac{{\widetilde {v'p'} }}{r} + \frac{1}{r}\widetilde {p'\left( {\frac{{\partial w'}}{{\partial \theta }} + v'} \right)}}_{\text{$V_w$}}  \underbrace{-\nu\left[ {\widetilde {{{\left( {\frac{{\partial w'}}{{\partial x}}} \right)}^2}}  + \widetilde {{{\left( {\frac{{\partial w'}}{{\partial r}}} \right)}^2}}  + \widetilde {\frac{1}{{{r^2}}}{{\left( {\frac{{\partial w'}}{{\partial \theta }} + v'} \right)}^2}} } \right]}_{\text{$\epsilon_w$}}
\end{split}
\end{equation}

\section*{Acknowledgments}
The authors gratefully acknowledge the financial support by the National Natural Science Foundation of China (Nos. 11772193, 42076210, 52122110, 61875123) and the National Youth Science Foundation of China (Nos. 52101322).

\section*{Declaration of interests}
The authors report no conflict of interest.

\bibliographystyle{jfm}
\bibliography{jfm-instructions}

\end{document}